\documentclass[aps,prd,11pt,notitlepage,longbibliography,nofootinbib,tightenlines,preprintnumbers]{revtex4-2}
\pdfoutput=1

\makeatletter
\renewcommand{\p@subsection}{}
\renewcommand{\p@subsubsection}{}
\makeatother

\usepackage{amsmath,amssymb,amsfonts}
\usepackage{graphicx}
\usepackage{color}
\usepackage{tikz}
\usepackage{dsfont}
\usepackage[pdftex]{hyperref}
\hypersetup{colorlinks=true, linkcolor=darkred, citecolor=blue, linktoc=page}
\definecolor{darkred}{rgb}{0.8,0.1,0.1}

\definecolor{3dcolor}{rgb}{0.96,0.89,0.76}
\definecolor{4dcolor}{rgb}{0.812,0.851,0.914}

\usepackage{MnSymbol}
\usepackage{subfigure}

\def\RR{{\mathds{R}}}
\def\ZZ{{\mathds{Z}}}

\newcommand{\trho}{\tilde\rho}
\newcommand{\vrho}{\varrho}

\DeclareMathOperator{\tr}{tr}

\def\Res{\mathop{\rm Res}}
\def\Im{\mathop{\rm Im}}
\def\Re{\mathop{\rm Re}}

\newcommand{\gym}{g_{\mathrm{YM}}}

\makeatletter\def\l@subsubsection#1#2{}%
\makeatother

\newcommand{\nocontentsline}[3]{}
\newcommand{\tocless}[2]{\bgroup\let\addcontentsline=\nocontentsline#1{#2}\egroup}

\newcommand{\be}{\begin{equation}}
	\newcommand{\ee}{\end{equation}}
\newcommand{\ba}{\begin{array}}
	\newcommand{\ea}{\end{array}}
\newcommand{\bea}{\begin{equation}\begin{aligned}}
		\newcommand{\eea}{\end{aligned}\end{equation}}

\newcommand{\mo}{\mathcal{O}}

\newcommand{\p}{\partial}

\newcommand{\deri}[2]{\frac{d #1}{d #2}}

\newcommand{\ehmo}[1]{\langle \hat{\mo}_{#1}\rangle}

\begin{document}

\title{One-point functions for doubly-holographic BCFTs and backreacting defects}

\author{Dongming He}
\email{dongming.he@vub.be}

\author{Christoph F.~Uhlemann} 
\email{christoph.uhlemann@vub.be}

\affiliation{Theoretische Natuurkunde, Vrije Universiteit Brussel and The International Solvay Institutes, Pleinlaan 2, B-1050 Brussels, Belgium}

\begin{abstract}
	We derive one-point functions in 4d $\mathcal N=4$ SYM with $\tfrac{1}{2}$-BPS boundaries, defects and interfaces which host large numbers of defect degrees of freedom. The theories are engineered by Gaiotto-Witten D3/D5/NS5 brane setups with large numbers of D5 and NS5 branes, and have holographic duals with fully backreacted 5-branes. 
	They include BCFTs with 3d SCFTs on the boundary which allow for the notion of double holography, as well as D3/D5 defects and interfaces with large numbers of D5-branes.
	Through a combination of supersymmetric localization and holography we derive one-point functions for 4d chiral primary operators.
\end{abstract}

\maketitle
\tableofcontents
\parskip 1mm

\section{Introduction}

Boundaries, defects and interfaces bring interesting dynamics to physical systems, often lead to qualitatively new phenomena, and have been studied extensively (a recent review is \cite{Andrei:2018die}). 
In string theory, boundaries hosting degrees of freedom which can support their own holographic dual allow for a notion of double holography, originating in \cite{Karch:2000ct,Karch:2000gx} and recently made precise in \cite{Uhlemann:2021nhu,Karch:2022rvr}. These setups involve large numbers of defect degrees of freedom which substantially backreact on the ambient theory.
In this work we focus on backreacting defects in $\mathcal N=4$ SYM, interfaces between $\mathcal N=4$ SYM theories with different gauge groups and couplings,  and boundary CFTs, all preserving half the supersymmetries and including theories relevant for double holography.
This broad class of theories is engineered by D3-branes ending on or intersecting groups of D5 and NS5 branes \cite{Gaiotto:2008sd,Gaiotto:2008sa,Gaiotto:2008ak}.

We focus on correlation functions of local operators. 
Unlike conventional CFTs, for defect, boundary and interface CFTs already the one-point functions of ambient operators are generally non-vanishing. With $x_3$ denoting the direction transverse to a planar defect, they take the form\footnote{If one views BCFTs as defined on $\rm AdS_4$ instead of flat space, and defect/interface CFTs on  two copies of $\rm AdS_4$, this turns into constant expectation values.}
\begin{align}\label{eq:one-point-R4-gen}
	\langle\mathcal O(x^\mu,x_3)\rangle&=\frac{a_{\mathcal O}}{x_3^{\Delta_{\mathcal O}}}~.
\end{align}
The coefficients $a_{\mathcal O}$ become meaningful once the normalization of the operators is fixed independently, e.g.\ by normalizing the two-point functions.
They contain dynamical information and are input for the boundary bootstrap \cite{Liendo:2012hy,Liendo:2016ymz}.
We study such one-point functions and collectively refer to them as defect one-point functions, whether the theory is a defect, interface or boundary CFT.

Defect one-point functions in $\mathcal N=4$ SYM have been studied extensively for D3/D5 defects and interfaces, and setups related by S-duality.
This includes holographic studies using probe branes in $\rm AdS_5\times S^5$, e.g.\ \cite{Nagasaki:2011ue,Nagasaki:2012re}, in which the number of D5-branes needs to be small compared to the number of D3-branes and extensive studies using integrability \cite{deLeeuw:2015hxa,Buhl-Mortensen:2015gfd,Buhl-Mortensen:2016pxs,deLeeuw:2017cop,deLeeuw:2019usb,DeLeeuw:2018cal,Linardopoulos:2025ypq}.
Probe branes and localization were combined in \cite{Robinson:2017sup}.
More recently, defect one-point functions for $\mathcal N=4$ SYM with a single D5-brane imposing Nahm pole boundary conditions (S-dual to setups with a single NS5) were obtained in \cite{Wang:2020seq,Komatsu:2020sup}, and these setups were also discussed in  \cite{Dedushenko:2020vgd,Dedushenko:2020yzd} and \cite{Beccaria:2022bjo}. 
One-point functions for BCFTs with less supersymmetry were discussed in \cite{Bason:2023bin}, with various low-rank examples.

Our primary interest here is in holographic theories engineered by D3, D5 and NS5 branes, but without the assumption of a probe or quenched limit. The defect, interface and boundary CFTs we consider are holographic, but the duals are far from $\rm AdS_5\times S^5$. 
Going beyond the probe limit is a natural generalization, e.g.\ for D3/D5, but one particular motivation is double holography. The latter needs setups with D5 and NS5 branes in large numbers, and ideally simultaneously, to realize 3d SCFTs on the boundary/defect which can support a holographic dual of their own, with a 3d free energy which is not bounded by the 4d central charge.
Generally, the backreaction of the 5-branes produces solutions of the form $\rm AdS_4\times S^2\times S^2\times\Sigma$ constructed in \cite{DHoker:2007zhm,DHoker:2007hhe,Aharony:2011yc,Assel:2011xz}.
These solutions underlie the double and wedge holography constructions in \cite{Uhlemann:2021nhu,Karch:2022rvr}, and have also featured in recent studies of supersymmetric indices \cite{Hatsuda:2024uwt}, SymTFTs \cite{GarciaEtxebarria:2024jfv} and the swampland \cite{Anastasi:2025puv}.

We derive one-point functions for ambient $\mathcal N=4$ SYM operators of the form $\tr (u\cdot \Phi)^J$ in a variety of setups.
This includes BCFTs with D3-branes ending on NS5 and D5 branes engineering 3d $\mathcal N=4$ quiver SCFTs, including the setups used for double holography in \cite{Uhlemann:2021nhu,Karch:2022rvr}.
In field theory terms these setups are governed by the long-quiver large-$N$ limit which was solved in localization in \cite{Uhlemann:2019ypp,Uhlemann:2020bek}.
We include D3/D5 defects and interfaces across which the gauge group is reduced by Nahm pole boundary conditions, where we allow the number of D5-branes to be large.
In the limit where $N_{\rm D5}\sqrt{\lambda}\ll N_{\rm D3}$ we recover results of probe brane calculations and connect to the results in \cite{Komatsu:2020sup} for a single D5. We further discuss S-dual interfaces engineered by D3-branes intersecting or ending on (large numbers of) NS5 branes. The sample of theories is selected such that it exhibits a multitude of S-duality relations, which we verify explicitly to provide extensive cross checks.

Our results are obtained using a combination of supersymmetric localization and AdS/CFT. We derived a precise relation between the saddle points dominating the matrix models and certain harmonic functions defining the supergravity duals in \cite{He:2024djr}, building on the holographic representation of Wilson loops associated with individual gauge nodes worked out in \cite{Coccia:2021lpp}. 
Through this relation the saddle points dominating the matrix models can be derived from the supergravity duals and vice versa.
We use this result to derive the saddle points from the supergravity duals, and then extract the one-point functions from the matrix models. It would be interesting to independently derive the one-point functions from the supergravity duals (as outlined in \cite[sec.~5]{Aharony:2011yc}), and, perhaps more ambitiously, to reproduce them from the intermediate description in double holography.

\subsection{Summary of results}

We consider 4d $\mathcal N=4$ SYM with $\tfrac{1}{2}$-BPS boundaries, defects or interfaces which preserve 3d $\mathcal N=4$ defect superconformal symmetry with an $SO(3)\times SO(3)$ R-symmetry. 
The 4d $\mathcal N=4$ vector multiplet splits into a 3d $\mathcal N=4$ vector multiplet and a hypermultiplet, schematically
\begin{align}\label{eq:4d-vector-split}
	(A_\mu,A_3,\Phi^I)+{\rm fermions}
	\quad &\rightarrow\quad
	(A_{\mu},\vec{Y})+{\rm fermions}
	\quad\cup\quad
	(A_3,\vec{X})+{\rm fermions}~,
\end{align}
where $x_3$ denotes the direction transverse to the defect/interface and $\mu=0,1,2$.
We study one-point functions of local chiral primary operators in the ambient 4d $\mathcal N=4$ SYM theories.
Their form, using the notation of \cite{Giombi:2009ds}, with a complex vector $u$ with $u^2=0$, is
\begin{align}\label{eq:2-pt-norm-sum}
	\mathcal O_J&= \tr\left(u\cdot \Phi\right)^J~,
	&
	\langle\mathcal O_J(x)\overline{\mathcal O}_{J'}(0)\rangle^{}_{\rm SYM}&=\frac{\delta_{JJ'}2^J}{|x|^{2\Delta}}\,,
\end{align}
where we fixed the normalization of $\mathcal O_J$ by specifying the two-point function in standard $\mathcal N=4$ SYM without boundaries or defects as in \cite{Komatsu:2020sup}.
Only operators which contain $SO(3)\times SO(3)$ singlets have non-trivial one-point functions, and this requires even $J$.
The one-point functions are computed through localization on $S^4$ for defect and interface CFTs  and on the hemisphere $HS^4$ for BCFTs, with the interface/boundary along an equatorial $S^3$.
The operators accessible through localization involve twisted-translations of the 3d hypermultiplet scalars $\vec{X}$ at the defect \cite{Giombi:2009ds,Wang:2020seq,Dedushenko:2020vgd}.
At $x^\mu=0$, $x_3=1$, corresponding to a pole of $(H)S^4$, they take the form $\tr(Y_1+iX_1)^J$, with $\vec{Y}$ the scalars appearing in half-BPS Wilson loops, and eq.~(\ref{eq:one-point-R4-gen}) becomes
$\langle\mathcal O(0,1)\rangle=a_{\mathcal O}$.

The conformal transformation to $(H)S^4$ mixes operators of different $J$ \cite{Gerchkovitz:2016gxx}, which has to be taken into account. This mixing is due to the appearance of local curvature invariants and independent of boundaries or defects. It can thus be accounted for using results for standard $\mathcal N=4$ SYM. Despite working at large $N$, this involves multitrace operators, as we explain in sec.~\ref{sec:one-point-functions}.
The results can be expressed concisely by first using the large-$N$ solution for the mixing problem among single-trace operators \cite{Rodriguez-Gomez:2016ijh,Rodriguez-Gomez:2016cem}, and then accounting for multitrace operators. This leads to
\begin{align}\label{eq:mixing-sum}
	\langle \mathcal O_{J=2}\rangle &= \langle \hat{\mathcal O}_{J=2}\rangle~,
	&
	\langle \mathcal O_{J=6}\rangle &= \langle \hat{\mathcal O}_{J=6}\rangle+\sqrt{\frac{3}{2}}\frac{1}{N}  \langle \hat{\mathcal O}_{J=2}\rangle^2~,
	\nonumber\\
	\langle \mathcal O_{J=4}\rangle &= \langle \hat{\mathcal O}_{J=4}\rangle~,
	&
	\langle \mathcal O_{J=8}\rangle &=  \langle \hat{\mathcal O}_{J=8}\rangle+\frac{4}{N}\langle \hat{\mathcal O}_{J=2}\rangle \langle \hat{\mathcal O}_{J=4}\rangle-\frac{1}{\sqrt{2}N}\langle \hat{\mathcal O}_{J=2}\rangle^2 ~,
\end{align}
where $\hat{\mathcal O}_J$ are `single-trace unmixed' auxiliary operators whose expectation values we give explicitly, and the relation to $\mathcal O_J$ can be extended to higher $J$ following the algorithm in app.~\ref{sec:app-diag}.

We discuss a set of examples which allows us to validate our results by explicitly verifying S-duality relations.
We start with the D3/D5 defect, which allows for many connections to existing literature: $N_{\rm D5}$ D5-branes intersecting $N_{\rm D3}$ D3-branes along a codimension-1 defect (fig.~\ref{fig:D3D5-defect}). The field theory is  4d $\mathcal N=4$ SYM with gauge group $U(N_{\rm D3})$ coupled to a set of $N_{\rm D5}$ 3d hypermultiplets localized on the defect. It was discussed in \cite{Aharony:2003qf,DeWolfe:2001pq}.
We find, for even $J$,
\begin{align}\label{eq:D3D5-defect-sum}
	\langle \hat{\mathcal O}_{J}\rangle&=
	\frac{N_{\rm D3}}{2^{J/2}\sqrt{J}} \left[
	\left(1+d^2\right) \, _2F_1\left(-\frac{J}{2},\frac{J}{2};2;d^2\right)
	+\left(1-d^2\right) \, _3F_2\left(\frac{1}{2},-\frac{J}{2},\frac{J}{2};\frac{3}{2},2;d^2\right)
	-\delta _{2,J}\right],
\end{align}
where
\begin{align}\label{eq:D3D5-sum-d-def}
	d&=\sqrt{1+\mathfrak N_{\rm D5}^2}-\mathfrak N_{\rm D5}~, & 
	\mathfrak N_{\rm D5}&=\frac{N_{\rm D5}\sqrt{\lambda}}{4\pi N_{\rm D3}}~.
\end{align}
The hypergeometric functions reduce to polynomials; example one-point functions obtained via (\ref{eq:mixing-sum}) are in (\ref{eq:D3D5-defect-ex}).
For $N_{\rm D5}=0$, the one-point functions vanish as expected. 
Expanding to linear order in $\mathfrak N_{\rm D5}\ll 1$ reproduces the holographic probe brane results of \cite{Nagasaki:2011ue}, providing a consistency check.

We then generalize the setup to allow for a number $k$ of D3-branes to end on the D5-branes (fig.~\ref{fig:D3D5-interface}), and for the couplings to differ on the two sides. 
This realizes interfaces between two 4d $\mathcal N=4$ SYM theories, one with gauge group $U(N)$ and coupling $g_R^2$ and one with gauge group $U(N+k)$ and coupling $g_L^2$, where the reduction of the gauge group at the interface is implemented by Nahm pole boundary conditions ((\ref{eq:D3D5-bc}), (\ref{eq:ti-decom-N5K})). 
We explicitly discuss the case of $N_{\rm D5}$ D5-branes with $k/N_{\rm D5}$ D3-branes ending on each D5.
For $k=0$ and $g_L=g_R$ this recovers the defect case.
We find, for the smaller-rank side with $N_{\rm D3}^{\rm R}=N$,
\begin{align}\label{eq:D3D5-gen-one-point-summary}
	\langle\hat{\mathcal O}_{J=2n}\rangle=
	\frac{N_{\rm D3}^{\rm R}}{2^n\sqrt{J}}
	\Bigg[-&\delta_{n,1}
	+
	\sum_{m=0}^{n}\frac{(-1)^m n\Gamma(m+n)}{\Gamma(1+m)\Gamma(2+m)\Gamma(1-m+n)}\left(\frac{(d^2-\kappa^2)\bar g^2}{g_R^2}\right)^{m}\Bigg\{2(d^2-\kappa^2)
	\nonumber\\
	&
	+\frac{(d+\kappa)(1-d^2+\kappa^2)}{d}\frac{1}{2m+1}\left(2m+2-\, _2F_1\left(1,m+\frac{1}{2};m+2;1-\frac{d^2}{\kappa^2}\right)\right)\Bigg\}\Bigg]\,,
\end{align}
where
\begin{align}\label{eq:D3D5-gen-one-point-summary-2}
	\mathfrak N_{\rm D5}&=\frac{N_{\rm D5}\sqrt{\bar g^2 N_{\rm D3}^{\rm R}}}{4\pi N_{\rm D3}^{\rm R}}~,
	&
	\bar g^2&=\frac{2g_L^2g_R^2}{g_L^2+g_R^2}~,
	&
	\kappa&=\frac{k}{4 N_{\rm D3}^{\rm R} \mathfrak N_{\rm D5}}~,
	&
	d&= \sqrt{(\kappa+\mathfrak N_{\rm D5})^2+1}-\mathfrak{N}_{\rm D5}~.
\end{align}
Example one-point functions for $g_L=g_R$ are in (\ref{eq:D3D5gen-ex}).
For $\mathfrak N_{\rm D5}\ll 1$ the leading-order expansion reproduces the strong-coupling results for interfaces with a single D5-brane in \cite{Komatsu:2020sup}, and $\kappa=0$ recovers the D5 defect case.
The generalization to multiple groups of D5-branes realizing general Nahm poles is outlined in sec.~\ref{sec:D3D5-gen}.
The S-dual NS5 interfaces are discussed below; this yields alternative expressions for the one-point functions which satisfy the expected S-duality relations.\footnote{For the D5 interfaces we only derive the matrix model saddle points from the gravity duals. For the S-dual NS5 interfaces we in addition independently verify the validity of the saddle points in the matrix models.}

We then turn to BCFTs in which 4d $\mathcal N=4$ SYM on a half space is coupled to a 3d $\mathcal N=4$ quiver SCFT on the boundary.
They are engineered by D3-branes ending on NS5-branes, with finite-length D3-branes suspended between the NS5-branes.
This engineers mixed 3d/4d quiver gauge theories where 4d $\mathcal N=4$ SYM couples to 3d SCFTs by gauging a global symmetry.
Additional D5-branes may add fundamental fields to the 3d quiver.
The BCFTs arise as strongly-coupled IR limits.
The first example is the brane setup in fig.~\ref{fig:D3NS5-branes}, with $N_5K$ D3-branes ending on $N_5$ NS5-branes, with $K$ D3-branes ending on each NS5. 
The quiver gauge theory is denoted as
\begin{align}\label{eq:D3NS5-quiver-sum}
	U(K)-U(2K)-\ldots - U((N_5-1)K)-\widehat{U(N_5K)}
\end{align}
where the node with hat denotes 4d $\mathcal N=4$ SYM on a half space and all other nodes are 3d gauge nodes localized on the boundary.
For $\langle \hat{\mathcal O}_J\rangle$ we find
\begin{align}\label{eq:D3NS5-OJ-sum}
	\langle \hat{\mathcal O}_{J}\rangle
	&=
	\frac{N_{\rm D3}}{(-2)^{J/2}\sqrt{J}}\left[\delta_{J,2}+2\Res_{u=-1}\left[ \frac{dx}{du}\frac{1-u}{1+u}\,T_J\left(\frac{x}{2\sqrt{k}}\right)\right]\right]\,,
	&
	k&=\frac{g_{\rm YM}^2K}{4N_5}~,
\end{align}
where $T_J(x)$ are the Chebyshev polynomials
and
\begin{align}\label{eq:x-def-sum}
	 x&=k\frac{1-u}{1+u}-\ln (-u)~.
\end{align}
The residues can be evaluated straightforwardly for any fixed $J$; examples for $\langle \mathcal O_J\rangle$ are in (\ref{eq:D3NS5-OJ-ex}).

We generalize this to BCFTs where the 3d SCFT on the boundary involves fundamental fields. 
The brane setup involves D3-branes ending on $N_{\rm D5}$ D5-branes and $N_{\rm NS5}$ NS5-branes, as illustrated in fig.~\ref{fig:brane-D5NS5-D3}.
In this case the numbers of 3d SCFT degrees of freedom and 4d degrees of freedom can be adjusted independently, and this type of theory was used for the implementation of double holography in \cite{Uhlemann:2021nhu,Karch:2022rvr}.
The quiver gauge theory with all 3d nodes balanced is 
\begin{align}\label{eq:D5NS5K-quiver-sum}
	U(R)-U(2R)-\ldots &- U(TR) - U(TR-Q)-\ldots - U(N_{\rm D3}+Q) - \widehat{U(N_{\rm D3})}
	\nonumber\\
	&\ \ \ \ \ \ \,\vert\\
	\nonumber & \ \ \ [N_{\rm D5}]	
\end{align}
where $N_{\rm D3}=RN_5+SN_{\rm D5}$ and $T=N_{5}+S$, $Q=N_{\rm D5}-R$.
We find
\begin{align}\label{eq:one-point-D3D5NS5-sum}
	\langle\hat{\mathcal O}_J\rangle&=
	\frac{N_{\rm D3}}{(-2)^{J/2}\sqrt{J}}\left[\delta_{J,2}
	+
	4\mathfrak{N}_5
	\Res_{u=-1}\left[\frac{dx}{du}
	\left(
	\mathfrak N_5k\frac{1-u}{1+u}
	-i \mathfrak{N}_{\rm D5}
	\ln\left(\frac{\frac{1-u}{1+u}-ie^\delta}{\frac{1-u}{1+u}+ie^\delta}\right)
	\right)
	T_J\left(\mathfrak{N}_5x\right)
	\right]
	\right]\,,
\end{align}
with  $x$ defined in (\ref{eq:x-def-sum}) and 
\begin{align}\label{eq:N5-def-sum}
	\mathfrak N_5&=\frac{N_5}{\sqrt{\lambda}}~, & 
	\mathfrak{N_{\rm D5}}&=\frac{N_{\rm D5}\sqrt{\lambda}}{4\pi N_{\rm D3}}~,
	&
	ke^\delta&=\frac{1-4 k\mathfrak{N}_5^2}{4\mathfrak{N}_5\mathfrak{N}_{\rm D5}}~,
	&
	\frac{\pi T}{\sqrt{\lambda}\mathfrak{N}_5}&=e^{\delta } k+
	2 \tan ^{-1}e^{\delta}~.
\end{align}
The last two equations determine the auxiliary parameters $k$ and $\delta$ in terms of the 5-brane numbers and the location of the flavors in the quiver, $T$.
Example one-point functions are in (\ref{eq:D3D5NS5-ex}), and for the special case $\delta=0$ in (\ref{eq:D3D5NS5-delta0-ex}) (this is the family of theories discussed in \cite[sec.~4]{Karch:2022rvr}).
S-duality maps this class of theories into itself and the one-point functions satisfy the expected relations, as we discuss in sec.~\ref{sec:D3D5NS5-BCFT}, where we also outline the extension to BCFTs with general balanced 3d quivers.

Finally, we discuss interfaces between 4d $\mathcal N=4$ SYM theories with independent ranks and gauge groups where the interfaces host 3d $\mathcal N=4$ SCFTs. We derive the saddle points for the matrix models for interfaces hosting general balanced 3d quiver SCFTs in sec.~\ref{sec:NS5-interface-saddle}. This extends the saddle points for BCFTs derived in \cite{He:2024djr} to interface CFTs. We also give general expressions for the one-point functions in (\ref{eq:one-point-NS5-int-gen}).
As special cases we explicitly discuss the S-duals of the aforementioned D3/D5 setups. We start with the dual of the D3/D5 defect, with brane setup given by replacing D5 by NS5 branes in fig.~\ref{fig:D3D5-defect}. The ICFT can be described by the quiver 
\begin{align}\label{eq:NS5-defect-quiver-sum}
	\widehat{U(N_{\rm D3})}-U(N_{\rm D3})-\ldots - U(N_{\rm D3})-\widehat{U(N_{\rm D3})}
\end{align}
For the auxiliary one-point functions we find
\begin{align}
	\langle\hat{\mathcal O}_J\rangle&=\frac{N_{\rm D3}}{(-2)^{J/2}\sqrt{J}}\left[\delta_{J,2}+2\mathfrak N_5 d \Res_{u=1}\left[ \frac{dv}{du} \left(  \frac{1-u}{1+u}-\frac{1+u}{1-u} \right)T_J\left(\mathfrak N_5(v-i\pi)\right)\right]\right]~,
\end{align}
where
\begin{align}\label{eq:D3NS5-d-def-sum}
	d&=\sqrt{1 + \mathfrak N_5^2} - \mathfrak N_5~,
	&
	\mathfrak N_5&=\frac{N_5}{\sqrt{\lambda}}~,
	&
	v&=\frac{d}{2\mathfrak N_5}\left(\frac{1-u}{1+u}+\frac{1+u}{1-u}\right)-\log u~.
\end{align}
The actual one-point functions $\langle\mathcal O_J\rangle$ are obtained from (\ref{eq:mixing-sum}); examples are in (\ref{eq:NS5-defect-one-point-ex}). The one-point functions $\langle O_J\rangle$ for D3/NS5 are related to those for the D3/D5 defect obtained from (\ref{eq:D3D5-defect-sum}) by replacing $\mathfrak{N}_5$ by $\mathfrak{N}_{\rm D5}$ defined in (\ref{eq:D3D5-sum-d-def}), which identifies $d$ between (\ref{eq:D3NS5-d-def-sum}) and (\ref{eq:D3D5-sum-d-def}), and including an overall factor $(-1)^{J/2}$. This is the expected S-duality relation (also discussed in \cite{Bason:2023bin}). We note that this relation does not hold at the level of the auxiliary one-point functions $\langle \hat O_J\rangle$.

As a more general special case we discuss interfaces with different gauge groups and couplings on the two sides, separated by a balanced 3d quiver SCFT with gauge groups of linearly increasing ranks.
The brane construction takes the form in fig.~\ref{fig:D3D5-interface}
and the quiver is
\begin{align}\label{eq:NS5-interface-quiver-sum}
	\widehat{U(N_{\rm D3}^L)}-U(N_1)-\ldots - U(N_{N_5-1})-\widehat{U(N_{\rm D3}^{\rm R})}
\end{align}
with the ranks of the 3d gauge groups decreasing in steps of $(N_{\rm D3}^L-N_{\rm D3}^{\rm R})/N_5$ from left to right.
The setup is S-dual to the aforementioned D3/D5 interfaces.
The auxiliary one-point functions are
\begin{align}
	\langle\hat{\mathcal O}_J\rangle&=\frac{N_{\rm D3}^{\rm R}}{(-2)^{J/2}\sqrt{J}}\biggr[\delta_{J,2}+ 2\mathfrak{N}_5\Res_{u=1}\left[\frac{dv}{du} \left(  (d-\kappa)\frac{1-u}{1+u}-(d+\kappa)\frac{1+u}{1-u}\right)T_J\left(\frac{N_5(v-i\pi)}{\sqrt{\lambda}}\right)\right]\biggr]~,
\end{align}
where
\begin{align}
	v&=\frac{g_R^2}{\bar g^2}\frac{d-\kappa}{2 \mathfrak N_5} \frac{1-u}{1+u}+\frac{g_L^2}{\bar g^2}\frac{d+\kappa}{2 \mathfrak N_5}\frac{1+u}{1-u}-\log u~,
    &
    \bar g^2&\equiv\frac{g_L^2 + g_R^2}{2}~,
\end{align}
and
\begin{align}
d &= \sqrt{(\kappa + \mathfrak N_5)^2 + 1} \;-\; \mathfrak N_5~, &
    \mathfrak N_5 &\equiv \frac{N_5}{\sqrt{\bar g^2 N_{\rm D3}^{\rm R}}}~,
	&
	\kappa &\equiv \frac{k}{4 N_{\rm D3}^{\rm R}\mathfrak N_5}~.
\end{align}
Examples for the one-point functions $\langle\mathcal O_J\rangle$ obtained from (\ref{eq:mixing-sum}) are in (\ref{eq: onepoin D3NS5gen}). 
These are related to the D3/D5 interface one-point functions resulting from (\ref{eq:D3D5-gen-one-point-summary}) in the expected way by S-duality. This provides an independent derivation of the one-point functions for the D3/D5 interfaces, with all steps verified independently in the matrix models, and a sensitive consistency check of our results.

The constructions and general formulas in sec.~\ref{sec:NS5-interface-saddle} more broadly cover ICFTs and BCFTs with general balanced 3d quiver SCFTs with flavor groups.
It would be interesting to generalize the explicit constructions to BCFTs and ICFTs with unbalanced 3d quivers, as used e.g.\ in \cite{Uhlemann:2023oea,Deddo:2023oxn}.

\section{One-point functions from localization}\label{sec:one-point-functions}

The one-point functions 
for the chiral primary operators in $\mathcal N=4$ SYM given in (\ref{eq:2-pt-norm-sum})
can be computed using supersymmetric localization on $S^4$ with the interface or boundary along an equatorial $S^3$ and the operators inserted at one of the poles.
Supersymmetric localization reduces the path integral to a (multi-)matrix integral, with one matrix integral for each ambient theory (e.g.\ one for a BCFT and one on each side of the interface for an ICFT). For a defect or boundary which hosts a 3d $\mathcal N=4$ SCFT which arises as strong-coupling limit of a 3d gauge theory an additional matrix integral arises for each 3d gauge node.
The combined matrix models take the schematic form
\begin{align}\label{eq:matrix-model}
	\mathcal Z&=\int \prod_{t=0}^L \prod_{i=1}^{N_t}da_i^{(t)} e^{-\mathcal F},
\end{align}
where $a_i^{(0)}$ and $a_i^{(L)}$ denote, respectively, the eigenvalues associated with the $\mathcal N=4$ SYM theories on the left and right half spaces, and the remaining $a_i^{(t)}$ denote eigenvalues associated with the 3d gauge nodes.
In suitable large-$N$ limits these integrals are dominated by saddle points, and we showed in \cite{He:2024djr} that these saddle points can be extracted from the holographic duals.
The one-point functions of $\mathcal O_J$ can be evaluated by including appropriate insertions in the matrix integral (\ref{eq:matrix-model}). 
One-point functions of the operators (\ref{eq:2-pt-norm-sum}) on $S^4$ are naively computed by insertions of monomials $\sum a_i^J$ into the matrix model, but this is modified due to mixing which we discuss now.

\subsection{Mixing and multi-traces}

As discussed in \cite{Gerchkovitz:2016gxx}, the conformal transformation from $\RR^4$ to $S^4$ leads to operator mixing of the form
\begin{align}
	\mathcal O_\Delta^{\RR^4} \rightarrow \mathcal O_{\Delta}^{S^4}+\frac{\alpha_1}{R^2} \mathcal O_{\Delta-2}+\frac{\alpha_2}{R^4} \mathcal O_{\Delta-2}+\ldots~,
\end{align}
where $R$ is the radius of $S^4$.
This mixing is a result of the conformal map and was related to contact terms and anomalies in \cite{Gerchkovitz:2016gxx}. It leads to mixing of two-point functions on $S^4$ which would not mix on $\RR^4$.
The procedure devised in \cite{Gerchkovitz:2016gxx} to account for this mixing is to Gram-Schmidt diagonalize the two-point functions on $S^4$ so that the resulting correlators can be mapped back to $\RR^4$.
This replaces the insertion of monomials into the matrix models with polynomials in the $a_i$.
This was worked out explicitly for $U(N)$ 4d $\mathcal N=4$ SYM in the `t Hooft large-$N$ limit in \cite{Rodriguez-Gomez:2016ijh,Rodriguez-Gomez:2016cem}.

Our interest is in 4d $\mathcal N=4$ SYM with boundaries, defects and interfaces. Such features affect the two-point functions: unlike in standard $\mathcal N=4$ SYM, the two-point functions in boundary and defect CFTs can be functions of an invariant cross ratio (e.g.~\cite[eq.~(2.7)]{Liendo:2012hy});
they contain dynamical information beyond coefficients, similar to 4-point functions in standard CFTs.
The mixing induced by the conformal transformation to $S^4$, however, results from the appearance of local curvature invariants and is expected to be independent of the boundary or defect.
Following \cite{Komatsu:2020sup,Beccaria:2022bjo,Bason:2023bin}, we apply the transformation which diagonalizes the two-point functions in standard $\mathcal N=4$ SYM without boundaries or defects also for our B/d/ICFTs. 

The Gram-Schmidt orthogonalization for standard $\mathcal N=4$ SYM at large $N$ has a closed-form solution in terms of Chebyshev polynomials $T_k$ \cite{Rodriguez-Gomez:2016ijh,Rodriguez-Gomez:2016cem}.
The insertions in this case would be
\begin{align}\label{eq:OJ-insert}
	\hat{\mathcal O}_J&=\left(\frac{4\pi i}{\sqrt{2\lambda}}\right)^J\frac{1}{\sqrt{J}}\sum_{i}f_J\left(a_i\right)\,,
	&
	f_J(a)&=\left(\frac{\sqrt{\lambda}}{4\pi}\right)^J\left[2T_J\left(\frac{2\pi a}{\sqrt{\lambda}}\right)+\delta_{J,2}\right]\,,
\end{align}
where $a_i$ are the eigenvalues associated with the corresponding 4d ambient theory in (\ref{eq:matrix-model}).
The Chebyshev polynomials in $f_J$ diagonalize the two-point functions in 4d $\mathcal N=4$ SYM with the normalization of \cite{Rodriguez-Gomez:2016ijh,Rodriguez-Gomez:2016cem}, 
and the overall factor realizes the normalization in (\ref{eq:2-pt-norm-sum}), as in \cite{Komatsu:2020sup}.

The above procedure ignores multi-trace operators, which are suppressed in standard $\mathcal N=4$ SYM. An argument can be made by writing the partition function in a way which exhibits manifest $N^2$ scaling and tracking factors of $N$ when taking derivatives to compute correlators. Each derivative is then accompanied by a factor $N^{-1}$, resulting in $\langle O_{n_1}O_{n_2}\ldots O_{n_m}\rangle\sim N^{2-m}$, where $O_{n_i}$ are single-trace operators (see e.g.\ \cite[app.~B]{Rodriguez-Gomez:2016ijh}).
This suppresses contributions of multi-trace operators in standard $\mathcal N=4$ SYM correlators at large $N$. However, genuinely non-trivial one-point functions $\langle O_n\rangle\sim N$ can overcome this suppression.
We therefore incorporate the finite-$N$ algorithm of \cite{Billo:2017glv}, which we review in app.~\ref{sec:app-diag}, and keep the leading large-$N$ contributions for all operators, including multi-traces. The results for the first operators can be expressed as
\begin{align}\label{eq:op-mixing}
	\mathcal O_{J=2}&=\hat{\mathcal O}_{J=2}~, & 
	\mathcal O_{J=6}&=\hat{\mathcal O}_{J=6}+\sqrt{\frac{3}{2}}\frac{1}{N}\left(\hat {\mathcal O}_{J=2}\right)^2~,
	\nonumber\\
	\mathcal O_{J=4}&=\hat{\mathcal O}_{J=4}~,
	&
	\mathcal O_{J=8}&=\hat{\mathcal O}_{J=8}+\frac{4}{N}\hat{\mathcal O}_{J=2}\hat{\mathcal O}_{J=4}-\frac{1}{\sqrt{2}N}\left(\hat {\mathcal O}_{J=2}\right)^2~,
\end{align}
and extension to higher $J$ is straightforward.
At leading order in $N$ the contributions of multi-trace operators to the one-point functions reduce to products of one-point functions, leading to (\ref{eq:mixing-sum}).
For defects treated as perturbative deformations, they do not affect the leading-order probe limit.

For the operators we consider, the matrix integrals with insertions are dominated by the unmodified saddle points (we assume $J$ is not of the same order as the free energy).
The computation of the one-point functions then reduces to evaluating the matrix model insertion on the unmodified saddle point.
With $\rho_{\rm 4d}$ denoting the eigenvalue density for the appropriate 4d ambient theory, 
\begin{align}\label{eq:one-point-int}
	\langle \hat{\mathcal O}_J\rangle&=\left(\frac{4\pi i}{\sqrt{2\lambda}}\right)^J\frac{1}{\sqrt{J}}\int da \rho_{\rm 4d}(a)
	 f_J(a)
	 =
	 \frac{1}{(-2)^{J/2}\sqrt{J}}\left[N\delta_{J,2}+2\int da\, \rho_{\rm 4d}(a)\,T_J\left(\frac{2\pi a}{\sqrt{\lambda}}\right)\right]~.
\end{align}
Since the $T_J(a)$ for odd $J$ are odd functions of $a$ while the eigenvalue densities are even, the expectation values are only non-zero for even $J$. This reflects that only one-point functions for even-$J$ operators are compatible with the $SO(3)\times SO(3)$ $R$-symmetry (see e.g.\ \cite{Wang:2020seq}).
The insertion can be spelled out explicitly using the representation of the Chebyshev polynomials as
\begin{align}
	T_n(x)
	&=\frac{n}{2}\sum_{m=0}^{\lfloor n/2\rfloor}(-1)^m\frac{\Gamma(n-m)}{\Gamma(1+m)\Gamma(1+n-2m)}(2x)^{n-2m}~.
\end{align}
With $J=2n$ the one-point function become (compare also \cite[(6.48),(6.49)]{Wang:2020seq})
\begin{align}\label{eq:one-point-gen}
	\langle\hat{\mathcal O}_{J=2n}\rangle
	&=
	\frac{1}{2^n\sqrt{2n}}
	\left[-N_{\rm D3}\delta_{n,1}
	+
	2n\sum_{m=0}^{n}\frac{(-1)^m\Gamma(m+n)}{\Gamma(1+2m)\Gamma(1-m+n)}\left(\frac{4\pi}{\sqrt{\lambda}}\right)^{2m}M_m
	\right]~,
\end{align}
where all theory-specific information is encoded in the moments of the eigenvalue density $\rho_{\rm{4d}}$,
\begin{align}\label{eq:Mn-def}
	M_m&=\int da\, a^{2m}\rho_{\rm 4d}(a)\,.
\end{align}

\subsection{Saddle points from supergravity}\label{sec:saddle-sugra-rel}
We will compute these one-point functions for concrete boundary, defect and interface CFTs from the saddle point eigenvalue densities. For a large class of BCFTs we derived the  saddle points in \cite{He:2024djr}. 
We also derived a general relation between the supergravity duals and the saddle points, which we will use to identify the saddle points for interface CFTs.

The supergravity solutions of \cite{DHoker:2007zhm,DHoker:2007hhe} which describe 4d $\mathcal N=4$ SYM with half-BPS boundaries, defects and interfaces have a warped product geometry $\rm AdS_4\times S^2\times S^2\times\Sigma$ and they ar entirely specified by a pair of harmonic functions $h_{1/2}$ on the Riemann surface $\Sigma$.
The full expressions for the 10d supergravity fields can be taken e.g.\ from the concise summary in \cite[sec.~4.1]{Coccia:2021lpp}. It is often convenient to express the harmonic functions and their duals in terms of their holomorphic and anti-holomorphic parts $\mathcal A_{1/2}$ as
\begin{align}
	h_1&=-i(\mathcal A_1-\bar{\mathcal A}_1)~, & 
	h_2&=\mathcal A_2+\bar{\mathcal A}_2~,
	\nonumber\\
	h_1^D&=\mathcal A_1+\bar{\mathcal A_1}~, & h_2^D&=i(\mathcal A_2+\bar{\mathcal A}_2)~.
\end{align}
The saddle point eigenvalue density can be derived in an implicit parametrization in terms of $h_{1/2}$ following \cite{He:2024djr}. Using that Wilson loops in antisymmetric representations are described by probe D5-branes along constant $h_2^D$, as shown in \cite{Coccia:2021lpp},
one finds the saddle point eigenvalue densities in an implicit parametrization as
\begin{align}\label{eq:rho-h12}
	h_2^D&={\rm const}~,&
	a&=\frac{h_2}{\pi \alpha'}~, &\rho&=\frac{2h_1}{\pi \alpha'}~,
\end{align}
where $a$ is the eigenvalue, $\rho$ the density at $a$ and the value of $h_2^D$ identifies the gauge node in the dual quiver gauge theory.
Here we will extend the explicit relations of \cite{He:2024djr} to interface CFTs.
For D3/D5 defects and NS5 interfaces we use this relation to derive the saddle point and verify it in field theory.
For D3/D5 interfaces we will use this relation to get the saddle points.

\section{D3/D5 defects and interfaces}\label{sec:D3D5}

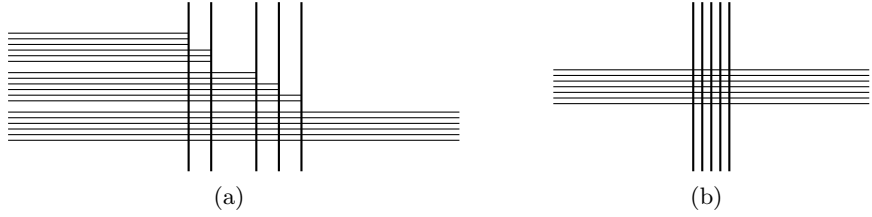
\begin{figure}
	\subfigure[][]{\label{fig:D3D5-interface}
	\begin{tikzpicture}[yscale=0.75,xscale=-1]
		\pgfmathsetmacro{\s}{0.3}
		\pgfmathsetmacro{\t}{0.1}
		
		\foreach \i in {5,...,7} \draw[thick] (\i*\s,-1.5) -- +(0,3);
		\foreach \i in {9,10} \draw[thick] (\i*\s,-1.5) -- +(0,3);
		
		\foreach \j in {-2,-1,0,1,2,3} \draw (-2*\s,-7.5*\t+\j*\t) -- (18*\s,-7.5*\t+\j*\t);
		
		\foreach \j in {5,6} \draw (5*\s,-7.5*\t+\j*\t) -- (18*\s,-7.5*\t+\j*\t);
		\foreach \j in {7,8} \draw (6*\s,-7.5*\t+\j*\t) -- (18*\s,-7.5*\t+\j*\t);
		\foreach \j in {9,10} \draw (7*\s,-7.5*\t+\j*\t) -- (18*\s,-7.5*\t+\j*\t);
		
		\foreach \j in {12,13,14} \draw (9*\s,-7.5*\t+\j*\t) -- (18*\s,-7.5*\t+\j*\t);
		\foreach \j in {15,16,17} \draw (10*\s,-7.5*\t+\j*\t) -- (18*\s,-7.5*\t+\j*\t);
	\end{tikzpicture}
	}\hskip 10mm
	\subfigure[][]{\label{fig:D3D5-defect}
		\begin{tikzpicture}[yscale=0.75]
			\pgfmathsetmacro{\s}{0.3}
			\pgfmathsetmacro{\t}{0.1}
			
			\foreach \i in {-2,...,2} \draw[thick] (0.4*\i*\s,-1.5) -- +(0,3);
			
			\foreach \j in {-3,-2,-1,0,1,2,3} \draw (-7*\s,\j*\t) -- (7*\s,\j*\t);
		\end{tikzpicture}
	}

	\caption{Left: Brane construction for an interface between $U(6)$ and $U(18)$ $\mathcal N=4$ SYM with Nahm pole boundary conditions. D3-branes are shown as as horizontal lines, D5-branes as vertical lines. 12 D3-branes on the left end on two groups of D5-branes to reduce the rank of the gauge group. The first 5-brane group contains 3 D5-branes with 2 D3-branes ending on each, the second contains 2 D5-branes with 3 D3-branes ending on each.
	Right: Defect in 4d $\mathcal N=4$ SYM realized by D5-branes intersecting D3-branes.\label{fig:D3D5-brane-gen}}
\end{figure}

We discuss defects, in which 4d $\mathcal N=4$ SYM with gauge group $U(N)$ is coupled to $N_{\rm D5}$ 3d hypermultiplets in the fundamental representation on a 3d defect, and interfaces.
The D3/D5 interface involves two 4d $\mathcal N=4$ SYM theories on half spaces, with gauge groups $U(N+k)$ and $U(N)$ with independent couplings, joined at an interface where boundary conditions reduce the gauge group from $U(N+k)$ to $U(N)$. 
The brane constructions are illustrated in fig.~\ref{fig:D3D5-brane-gen}.

Terminating D3-branes on D5-branes imposes a Nahm pole boundary condition. For the 4d fields split as in (\ref{eq:4d-vector-split}),
\begin{align}\label{eq:D3D5-bc-0}
	F_{\mu\nu}\big\vert_{x_3=0}&=0~, &
	Y_i\big\vert_{x_3=0}&=0~,
	&
	D_3 X_i -\frac{i}{2}\epsilon_{ijk}[X_j,X_k]\big\vert_{x_3=0}&=0~.
\end{align}
The general solution for $\vec{X}$ involves a pole specified by a set of $SU(2)$ generators $t^i$,
\begin{align}\label{eq:D3D5-bc}
	X^i&\sim\frac{t^i}{x^3}~,
	&
	[t^i,t^j]&=i\epsilon^{ijk}t^k~.
\end{align}
How the D3-branes end on the D5-branes determines the representation of $t_i$.
For the interface, we denote the fields on the larger-rank side by superscript $+$ and  those on the smaller-rank side by a superscript $-$, as in \cite[(2.18)]{Komatsu:2020sup}.
At the interface
\begin{align}
	A_\mu^+&=\begin{pmatrix} A_\mu^- & \star \\ \star & \star \end{pmatrix}\,,
	&
	\vec{Y}^+&=\begin{pmatrix} \vec{Y}^- & \star \\ \star & \star \end{pmatrix}\,,
	&
	\vec{X}^+&=\begin{pmatrix} \vec{X}^- & \star \\ \star & \frac{1}{x_3}\vec{t} \end{pmatrix}\,.
\end{align}
The $U(N)$ part is continuous while the $k\times k$ part satisfies the Nahm pole boundary condition.
For an interface with $k=N_{D5} K$ D3-branes ending on $N_{D5}$ D5-branes, with $K$ D3 ending on each D5, the $t_i$ are block-diagonal $k\,{\times}\, k$ matrices with $K\,{\times}\, K$ blocks $t_i^{K\times K}$ on the diagonal,
\begin{align}\label{eq:ti-decom-N5K}
	t_i&=t_i^{K\times K}\oplus \ldots \oplus t_i^{K\times K}~,
\end{align}
with $t_i^{K\times K}$ denoting the $K$-dimensional irreducible representation.
A global symmetry emerges \cite[sec.~2.4]{Gaiotto:2008sa}, which arises in the brane setup from the $U(N_{D5})$ associated with the $N_{\rm D5}$ D5 branes. 
For $N_{\rm D5}=1$ this setup was considered in \cite{Komatsu:2020sup}.
We will mainly focus on this setup with general $N_{\rm D5}$.

The generalization to multiple groups of D5-branes with the same number of D3-branes ending on each D5-brane within a group is straightforward and will be outlined.

\subsection{Supergravity duals}

The supergravity dual for two 4d $\mathcal N=4$ SYM theories on half spaces, with independent gauge couplings, joined at a defect or interface with $N_{\rm D5}$ D5-branes such that all D5-branes have the same numbers of D3-branes ending on them (zero for a defect) can be constructed from the solutions of \cite{DHoker:2007zhm,DHoker:2007hhe}.
It is given by a particular pair of holomorphic functions, namely
\begin{align}\label{eq:D3D5-cA12}
	\mathcal A_1&=\frac{\pi \alpha'}{4}\left(K_0 e^z-K_1 e^{-z}\right)-\frac{i\alpha'}{4}N_{\rm D5}\ln\tanh\left(\frac{i\pi}{4}-\frac{z}{2}\right)~,
	\nonumber\\
	\mathcal A_2&=\frac{\pi\alpha'}{4}\left(K_2 e^z+K_3e^{-z}\right)~.
\end{align}
The full expressions for the 10d supergravity fields can be taken e.g.\ from the concise summary in \cite[sec.~4.1]{Coccia:2021lpp}.
The 4 parameters $K_i$ determine the 2 gauge couplings and the numbers of semi-infinite D3-branes on the left/right side of the interface as follows,
\begin{align}
	e^{2\phi_L}&=\frac{K_3}{K_1}~, 
	&N_{\rm D3}^{\rm L}&=\frac{\pi}{2} (K_0K_3+K_1K_2)+K_3 N_{\rm D5}~,
	\nonumber\\
	e^{2\phi_R}&=\frac{K_2}{K_0}~,&
	k\equiv N_{\rm D3}^{\rm L}-N_{\rm D3}^{\rm R}&=N_{\rm D5}(K_3-K_2)~.
\end{align}
We emphasize that $N_{\rm D5}$ does not need to be small.
The gauge couplings on the two sides are given by (noting the dilaton convention $\tau=\chi+i e^{-2\phi}$)
\begin{align}
	e^{2\phi_{L/R}}&=\frac{g_{L/R}^2}{4\pi}\,.
\end{align}
For given supergravity parameters $(K_{0,1,2,3},N_{\rm D5})$ we can read off the field theory parameters $(g_{L/R},N_{\rm D3}^{L/R},N_{\rm D5})$.
We will also need the inverse relation. For later convenience we introduce
\begin{align}\label{eq:cND5-kappa-def}
	\mathfrak N_{\rm D5}&\equiv\frac{N_{\rm D5}\sqrt{\bar g^2N_{\rm D3}^{\rm R}}}{4\pi N_{\rm D3}^{\rm R}}\,,
	&
	\kappa&\equiv\frac{k}{4N_{\rm D3}^{\rm R}\mathfrak{N}_{\rm D5}}\,,
	&
	d&\equiv \sqrt{(\kappa+\mathfrak N_{\rm D5})^2+1}-\mathfrak{N}_{\rm D5}\,,
	&
	\bar g^2&\equiv\frac{2g_L^2g_R^2}{g_L^2+g_R^2}\,,
\end{align}
where we parametrize the number of D5-branes by $\mathfrak N$ and the difference in ranks by $\kappa$.
We then find as solution with positive $K_i$
\begin{align}
	K_0&=\frac{4\pi}{g_R^2}K_2\,,
	&
	K_2&=\frac{1}{2\pi}\left(d-\kappa\right) \sqrt{\bar g^2N_{\rm D3}^{\rm R}}\,,
	\nonumber\\
	K_1&=\frac{4\pi}{g_L^2}K_3\,,
	&
	K_3&=\frac{1}{2\pi}\left(d+\kappa\right) \sqrt{\bar g^2N_{\rm D3}^{\rm R}}\,.
\end{align}
We further define combinations which will play prominent roles as
\begin{align}\label{eq:D3D5-Khat-c-def}
	\hat K^2&\equiv K_2K_3=
	\frac{\bar g^2N_{\rm D3}^{\rm R}}{4\pi^2}
	\left(d^2-\kappa ^2\right)\,,
	&
	c&\equiv \frac{K_2+K_3}{2\hat K}=\frac{d}{\sqrt{d^2-\kappa^2}}\,.
\end{align}
We note that the limit $K_{0,2}\rightarrow 0$ describes a BCFT, and was discussed e.g.\ in \cite{Chaney:2024bgx}.

\subsection{Eigenvalue densities}\label{sec:D3D5-densities}

The saddle point eigenvalue densities can be derived from the supergravity duals through the identification of Wilson loop expectation values between the two descriptions, following \cite{He:2024djr}.
The D3/D5 defect and interface CFTs both host a single family of Wilson loops localized on the defect in antisymmetric representations:
For the D3/D5 interface the Nahm pole b.c.\ set part of the gauge field to zero on the interface, reducing $U(N+k)$ to $U(N)$, while the remaining components of $A$ are continuous across the interface. The same applies for the corresponding components of $\vec{Y}$. The limits of taking a Wilson loop to the interface from the left or from the right then agree, and both are determined by the $U(N)$ eigenvalues.

To extract the eigenvalues we need the holographic representation of antisymmetric Wilson loops.
The D5$'$-branes representing Wilson loops are embedded along curves in $\Sigma$ with $h_2^D={\rm const}$ \cite{Coccia:2021lpp}. 
For the D3/D5 setups the only embedding localized between the $\rm AdS_5\times S^5$ regions is
\begin{align}\label{eq:D3D5-Wilson}
	{\rm D5}': \qquad\Re(z)=\frac{1}{2}\ln\frac{K_3}{K_2}~.
\end{align}
As expected, we get one defect Wilson loop and access to one eigenvalue density. 
In the BCFT limit $K_{0,2}\rightarrow 0$ the Wilson loop disappears: the gauge field is then set to zero on the boundary. A matrix model description can still be extracted from the S-dual description, which we do in sec.~\ref{sec:D3NS5}.

\begin{figure}
	\includegraphics[width=0.36\linewidth]{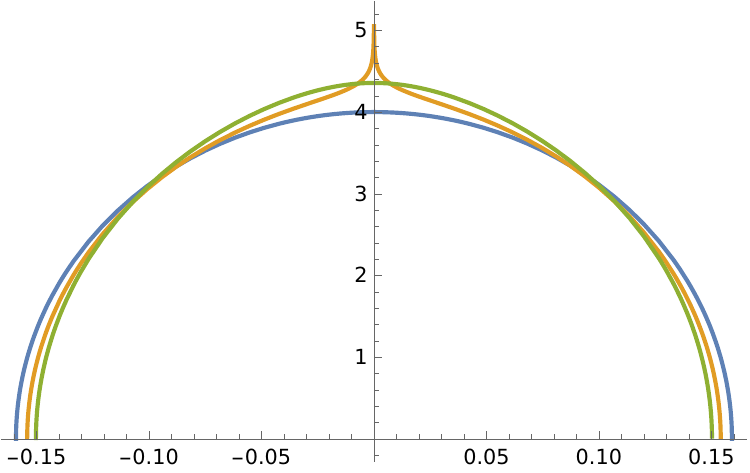}
	\caption{Eigenvalue densities for the D3/D5 defect with $k=0$ (yellow) and interface with $k\neq0$ (green) compared to unmodified $\mathcal N=4$ SYM (blue) at the same $N$ and $g$.\label{fig:D3D5-defect-density}}
\end{figure}

The saddle point eigenvalue density can be derived in an implicit parametrization in terms of $h_{1/2}$ from (\ref{eq:rho-h12}), with $h_2^D={\rm const}$ solved by (\ref{eq:D3D5-Wilson}).
Upon solving for $a$ we obtain an explicit form for $\rho(a)$ which is 
a deformation of the familiar Wigner semi-circle distribution
\begin{align}\label{eq:D3D5-defect-rho}
	\rho(a)&=
	\frac{K_0K_3+K_1K_2}{K_2K_3}\sqrt{K_2K_3-a^2}-\frac{N_{\rm D5}}{\pi }
	\ln \left(\frac{K_2+K_3-2\sqrt{K_2K_3-a^2}}{\sqrt{4a^2+(K_2-K_3)^2}}\right).
\end{align}
It has compact support with an (integrable) singularity at $a=0$ if $K_2=K_3$, otherwise it is regular at $a=0$.
The Wigner semi-circle is recovered for $N_{\rm D5}=0$.
Note that this captures the Janus solution with merely different gauge couplings as special case. It also captures the defect case without Nahm pole.
We note that $N_{\rm D5}$ enters the relation between $K_i$ and $N_{\rm D3}^{L/R}$, so both terms are modified compared to standard $\mathcal N=4$ SYM.
We can express this more directly in terms of field theory quantities using (\ref{eq:D3D5-Khat-c-def}) as
\begin{align}\label{eq:rho-D3-D5-1}
	\rho(a)&=
\frac{8\pi}{\bar g^2}\sqrt{\hat K^2-a^2}-\frac{N_{\rm D5}}{2\pi}
\ln \left(\frac{\hat K c-\sqrt{\hat K^2-a^2}}{\hat K c+\sqrt{\hat K^2-a^2}}\right),
\end{align}
Plots are in fig.~\ref{fig:D3D5-defect-density}. For the special case of a D3/D5 defect with equal 4d gauge groups on both sides the matrix model will be discussed in detail below.
The moments of the densities are given by
\begin{align}\label{eq:D3D5-Mm}
	M_m&=\hat K^{2 m+1}
	\frac{ \Gamma \left(m+\frac{1}{2}\right)}{2 \sqrt{\pi } \Gamma (m+2)}
	\left(\frac{8 \pi^2\hat K}{\bar g^2}+\frac{1}{c}N_{\rm D5}+N_{\rm D5}\frac{1-\, _2F_1\left(1,m+\frac{1}{2};m+2;\frac{1}{1-c^2}\right)}{c(2 m+1)}\right)\,.
\end{align}
The hypergeometric functions reduce to polynomials for non-negative integer $m$ and 
for $m=0$ we recover the normalization
\begin{align}
	\int_{|a|<\hat K} da\,\rho(a)&=\min\left\lbrace N_{\rm D3}^L,N_{\rm D3}^{\rm R}\right\rbrace\,.
\end{align}
This is in line with the initial discussion and the expectation that the Wilson loops give access to the eigenvalue densities on the smaller-rank side of the interface when $N_{\rm D3}^L\neq N_{\rm D3}^{\rm R}$.

We note in passing that the expectation value of the fundamental Wilson loop localized on the defect/interface is
\begin{align}
	\langle W_f\rangle&=\frac{1}{N}\sum_{i=1}^N e^{2\pi a}\Big\vert_{\rm saddle}
	&&\rightarrow&
	\ln\langle W_f\rangle=2\pi \hat K~.
\end{align}
with $\hat K$ expressed in terms of field theory data in (\ref{eq:D3D5-Khat-c-def}).
The expectation values for antisymmetric Wilson loops can be obtained straightforwardly from the above results.

\subsection{D3/D5 defect}\label{sec:D3D5-defect}

We start with the special case of defects in 4d $\mathcal N=4$ SYM with gauge group $U(N)$ (fig.~\ref{fig:D3D5-defect}), i.e.\ the same gauge groups and couplings on both sides. 
The D5-branes add $N_{\rm D5}$ 3d hypermultiplets.

\textbf{Matrix Model:}
The matrix model appeared first in \cite{Robinson:2017sup}, where solutions were discussed in the quenched limit, and it was derived in \cite{Wang:2020seq}. It was analyzed in \cite{Komatsu:2020sup} for $N_{\rm D5}=1$ with arbitrary $\lambda$ and in \cite{Beccaria:2022bjo} including $1/N$ corrections.
Here we allow for large D5-brane numbers, of the same order as $N$, but assume large $N$ and $\lambda$.
The matrix model is
\begin{align}
	\mathcal Z&=\int [da]\Delta(a)^2\exp\left\lbrace -\frac{8\pi^2 N}{\lambda}\sum_{i=1}^Na_i^2\right\rbrace\prod_{j=1}^N\left(2\cosh(\pi a_j)\right)^{-N_f},
\end{align}
where $\lambda =Ng_{\rm YM}^2$ and the integration measure and Vandermonde determinant are
\begin{align}
	[da]&=\prod_{i=1}^N da_i~,
	&
	\Delta(a)&=\prod_{i<j}(a_i-a_j)~.
\end{align}
We interchangeably use $N_{\rm D5}$ and $N_f$ to denote the number of fundamental 3d hypermultiplets.
Collecting the integrand in an exponential leads to
\begin{align}
	\mathcal Z&=\int [da]e^{-\mathcal F}~,
	&
	\mathcal F&=-\ln\Delta(a)^2+\frac{8\pi^2 N}{\lambda}\sum_{i=1}^Na_i^2
	+N_f\sum_{j=1}^N\ln\left(2\cosh(\pi a_j)\right)~,
\end{align}
where $\ln\Delta(a)^2=\sum_{i\neq j}\ln|a_i-a_j|$.
Upon introducing an eigenvalue density normalized such that $\int da \rho(a)=N$, this becomes
\begin{align}\label{eq:D3D5-defect-cF}
	\mathcal F&=-\int da da' \rho(a)\rho(a')\ln|a-a'|+
	\frac{8\pi^2 N}{\lambda}\int da \rho(a) a^2+N_f \int da \rho(a) \ln(2\cosh(\pi a))~.
\end{align}
The saddle point equation, obtained by varying $\rho$ while including a Lagrange multiplier $C_0$ to enforce the correct normalization, reads
\begin{align}\label{eq:D3D5-defect-saddle-0}
	-2\int da'\rho(a')\ln|a-a'|+\frac{8\pi^2 N}{\lambda}a^2+N_f\ln(2\cosh(\pi a))
	&=-2C_0~.
\end{align}

The solution is a special case of (\ref{eq:D3D5-defect-rho}).
For the defect we have $K_3=K_2$ and $K_1=K_0$ to get the same ranks for the gauge groups to either side and the same couplings. This implies
\begin{align}\label{eq:cKhd-defect}
	c&=1~, & \hat K&=\frac{\sqrt{\lambda}}{2\pi}d~, & d&=\sqrt{1+\mathfrak N_{\rm D5}^2}-\mathfrak N_{\rm D5}~.
\end{align}
The saddle point becomes
\begin{align}
	\rho(a)&=\frac{8\pi N}{\lambda}
	\sqrt{1-a^2/\hat K^2}-\frac{N_{\rm D5}}{2\pi }
	\ln \left(\frac{1-\sqrt{1-a^2/\hat K^2}}{1+\sqrt{1-a^2/\hat K^2}}\right).
\end{align}
When $\hat K$ is large, the overall scale of the eigenvalues $a$ is large and $\ln(2\cosh(\pi a))\approx \pi |a|$. 
The saddle point equation (\ref{eq:D3D5-defect-saddle-0}) is then satisfied with
\begin{align}
	C_0&=\int da'\rho(a')\ln|a'|
	=
	\frac{2 \pi ^2 N}{\lambda}\hat K^2\left(2\ln\left(\frac{\hat K}{2}\right)-1\right)
	+\hat K N_{\rm D5}\left(\ln\left(\frac{\hat K}{2}\right)-1\right)~.
\end{align}

As a cross check we compute the defect free energy in the matrix model, starting from (\ref{eq:D3D5-defect-cF}).
Using the saddle point equation (\ref{eq:D3D5-defect-saddle-0}) with $\ln(2\cosh(\pi a))\sim |\pi a|$ leads to
\begin{align}\label{eq:D3D5-defect-cF-on-shell}
	\mathcal F&=\frac{4\pi^2 N}{\lambda}\int da \rho(a) a^2+\frac{N_f}{2} \int da \rho(a) \pi |a|-N C_0~.
\end{align}
Evaluating this explicitly leads to
\begin{align}
	\mathcal F&=\frac{\pi^2 N}{3\lambda} K_2^3 (3 \pi K_0+2N_{\rm D5})
	+\frac{N_f}{2} \int da \rho(a) |\pi a|-N C_0
	\nonumber\\
	&=N_{\rm D3}^2\left(\frac{3}{4}-\frac{1}{2}\ln\frac{\lambda}{16\pi^2}\right)+\frac{1}{12} N_{\rm D3}^2 \left(9-d^4-8 d^2-12\ln d\right)~.
\end{align}
The first part is the free energy of $\mathcal N=4$ SYM without defect, the second is the defect free energy.
This defect free energy matches the supergravity computation of the defect entropy in \cite[(3.36)]{Estes:2014hka}.

\textbf{Defect one-point functions:} 
The moments (\ref{eq:D3D5-Mm}) simplify for the defect case. The hypergeometric functions can be expanded and we find
\begin{align}
	M_m&=\hat K^{2 m+1}\frac{\Gamma \left(m+\frac{1}{2}\right)}{2 \sqrt{\pi } \Gamma (m+2)}
	\left(\frac{8 \pi ^2 N}{\lambda}\hat K+N_{\rm D5}+\frac{N_{\rm D5}}{2 m+1}\right)~.
\end{align}
The auxiliary one-point functions (\ref{eq:one-point-gen}) become
\begin{align}
	\langle\hat {\mathcal O}_{J=2n}\rangle
	=
	\frac{1}{2^n\sqrt{J}}
	\Bigg[-N\delta_{n,1}
	+
	\frac{n\hat K}{\sqrt{\pi}} \sum_{m=0}^{n}
	&\,
	\frac{(-1)^m\Gamma(m+n)}{\Gamma(1+2m)\Gamma(1-m+n)}\left(\frac{4\pi\hat K}{\sqrt{\lambda}}\right)^{2m}
	\times\nonumber\\&
	\frac{\Gamma \left(m+\frac{1}{2}\right)}{\Gamma (m+2)}
	\left(\frac{8 \pi ^2 N}{\lambda}\hat K+N_{\rm D5}+\frac{N_{\rm D5}}{2 m+1}\right)
	\Bigg]\,.
\end{align}
For the last term in the parenthesis the $1/(2m+1)$ combines with $\Gamma(1+2m)$ in the denominator and we can use the following identity
\begin{align}\label{eq:sum-hyp-id}
	F(x)&=\frac{2n}{\sqrt{\pi}}
	\sum_{p=0}^n(-1)^p\frac{\Gamma(n+p)\Gamma(p+1/2)}{\Gamma(1+n-p)\Gamma(2p+2)\Gamma(p+2)}x^p
	=
	2\,{}_3F_2\left(\frac{1}{2},-n,n;\frac{3}{2},2;\frac{x}{4}\right)~,
\end{align}
where ${}_pF_q$ is the generalized hypergeometric function. 
For the remaining terms we use 
\begin{align}
	x F'(x)&=\, _2F_1\left(-n,n;2;\frac{x}{4}\right)-\, _3F_2\left(\frac{1}{2},-n,n;\frac{3}{2},2;\frac{x}{4}\right)~.
\end{align}
Altogether this yields the auxiliary one-point functions as
\begin{align}\label{eq:D3D5-defect-one-point-1}
	\langle \hat{\mathcal O}_{J=2n}\rangle&=
	\frac{N_{\rm D3}}{2^n\sqrt{J}} \left[
	\left(d^2+1\right) \, _2F_1\left(-n,n;2;d^2\right)
	-\delta _{1,n}
	-\left(d^2-1\right) \, _3F_2\left(\frac{1}{2},-n,n;\frac{3}{2},2;d^2\right)\right]
\end{align}
with $d$ for the defect given in (\ref{eq:cKhd-defect}), where we note that $0<d<1$.
The actual one-point functions are obtained from (\ref{eq:mixing-sum}) and the first examples are
\begin{align}\label{eq:D3D5-defect-ex}
	\langle \mathcal O_{J=2}\rangle&=
	-\frac{d^4+2 d^2-3}{6 \sqrt{2}}N_{\rm D3}~,
	\qquad\qquad\qquad\qquad
	\langle \mathcal O_{J=4}\rangle=
	\frac{6 d^6-d^4-20 d^2+15}{60}N_{\rm D3}~,
	\nonumber\\
	\langle \mathcal O_{J=6}\rangle&=-\frac{190 d^8-344 d^6-371 d^4+1050 d^2-525}{840 \sqrt{6}}N_{\rm D3}~,
		\nonumber\\
	\langle \mathcal O_{J=8}\rangle&=\frac{644 d^{10}-2161 d^8+880 d^6+4312 d^4-5880 d^2+2205}{5040 \sqrt{2}}N_{\rm D3}~.
\end{align}
We will derive alternative expressions for the same one-point functions from the S-dual setups in sec.~\ref{sec:NS5-defect}.
Upon expanding the one-point functions to linear order in $N_{\rm D5}$ they simplify and reproduce the results for $N_{\rm D5}=1$ quoted in \cite[(3.49)]{Komatsu:2020sup}.

\subsection{D3/D5 and Janus interfaces}\label{sec:D3D5-gen}

We now discuss interfaces with partial Nahm pole boundary conditions and independent gauge groups and couplings on the two sides, i.e.\ $k\neq 0$ and $g_L$ and $g_R$ independent. 
The one-point functions on the two sides then in general differ.
When the ranks of the gauge groups differ, the relation between saddle points and supergravity solutions of sec.~\ref{sec:saddle-sugra-rel} gives access to the saddle point eigenvalues and thus one-point functions on the side with smaller rank, as discussed in sec.~\ref{sec:D3D5-densities}. The S-dual setups discussed in sec.~\ref{sec:NS5-interface-one-pt}, however, provide the one-point functions on both sides. Perhaps unsurprisingly, the results on the two sides are related by continuation to negative $k$.

The simplest ICFT is two 4d $\mathcal N=4$ SYM theories with the same gauge groups but different couplings joined at an interface with no additional degrees of freedom. This means $N_{\rm D5}=k=0$ and $N_{\rm D3}^L=N_{\rm D3}^{\rm R}=N_{\rm D3}$.
The eigenvalue density (\ref{eq:rho-D3-D5-1}) reduces to a Wigner semi-circle, 
\begin{align}
		\rho(a)&=
	\frac{8\pi}{\bar g^2}\sqrt{\hat K^2-a^2}~, & \hat K&=\frac{\bar g^2N_{\rm D3}}{4\pi^2}~,
\end{align}
with $\bar g$ defined in (\ref{eq:cND5-kappa-def}) a deformed coupling compared to standard $\mathcal N=4$ SYM.\footnote{When combining two partition functions for $\mathcal N=4$ SYM on half spaces the classical action parts in the matrix models add as $4\pi^2g_L^{-2}\sum_i a_{i,L}^2+4\pi^2g_R^{-2}\sum_i a_{i,R}^2=8\pi^2\bar g^{-2}\sum_i a_i^2$.}
The combination $\bar g$ is symmetric under exchange of $g_L$ and $g_R$.
The moments are
\begin{align}
	M_m&=	\frac{8 \pi^2}{\bar g^2} \hat K^{2 m+2}
	\frac{ \Gamma \left(m+\frac{1}{2}\right)}{2 \sqrt{\pi } \Gamma (m+2)}~.
\end{align}
The auxiliary one-point functions (\ref{eq:one-point-gen}) become
\begin{align}
	\langle\hat{\mathcal O}_{J=2n}\rangle
	&=
	\frac{1}{2^n\sqrt{J}}
	\left[-N_{\rm D3}\delta_{n,1}
	+
	\sum_{m=0}^{n}\frac{(-1)^m2n\Gamma(m+n)}{\Gamma(1+2m)\Gamma(1-m+n)}\left(\frac{4\pi}{\sqrt{\lambda_{L/R}}}\right)^{2m}M_m
	\right]
	\nonumber\\
	&=
	\frac{N_{\rm D3}}{2^n\sqrt{2n}}\left(2 \, _2F_1\left(-n,n;2;\frac{\bar g^2}{g_{L/R}^2}\right)-\delta _{1,n}\right)\,,
\end{align}
where $g_{L/R}$ originate in the argument of the Chebyshev polynomial in (\ref{eq:OJ-insert}) depending on which side the one-point function is computed on.
For standard $\mathcal N=4$ SYM with $\bar g=g_R=g_L$ the argument of the hypergeometric function becomes one and the one-point functions vanish, as expected.

For the general case we start from the auxiliary one-point functions expressed as in (\ref{eq:one-point-gen}) with the moments in (\ref{eq:D3D5-Mm}).
Altogether, we find
\begin{align}\label{eq:D3D5-gen-one-point-2}
	\langle\hat{\mathcal O}_{J=2n}\rangle&=
	\frac{1}{2^n\sqrt{J}}
	\left[-N_{\rm D3}^{\rm R}\delta_{n,1}
	+
	\sum_{m=0}^{n}\frac{(-1)^m n\Gamma(m+n)}{\Gamma(1+m)\Gamma(2+m)\Gamma(1-m+n)}\left(\frac{2\pi \hat K}{\sqrt{\lambda_{\rm R}}}\right)^{2m}\tilde F(m)
	\right]\,,
	\nonumber\\
	\tilde F(m)&=\frac{8 \pi^2\hat K^2}{\bar g^2}+\frac{\hat K N_{\rm D5}}{c}\frac{1}{2m+1}\left(2m+2-\, _2F_1\left(1,m+\frac{1}{2};m+2;\frac{1}{1-c^2}\right)\right)~.
\end{align}
These are the one-point functions for the lower-rank side, to which $\lambda_{\rm R}$ and $N_{\rm D3}^{\rm R}$ refer, for interfaces with independent ranks and couplings on the two sides and an arbitrary number of D5-branes corresponding to Nahm poles of the form (\ref{eq:ti-decom-N5K}).
Here we have
\begin{align}
	\frac{\hat KN_{\rm D5}}{c}&=\frac{(d+\kappa)(1-d^2+\kappa^2)}{d} N_{\rm D3}^{\rm R}~,
	&
	\frac{8\pi^2\hat K^2}{\bar g^2}&=2(d^2-\kappa^2)N_{\rm D3}^{\rm R}~, & \frac{1}{1-c^2}&=1-\frac{d^2}{\kappa^2}~.
\end{align}
This expresses all quantities in terms of $d$, $\kappa$, $\mathfrak N_{\rm D5}$, $\bar g^2$,
which were defined in (\ref{eq:cND5-kappa-def}), and $N_{\rm D3}^{\rm R}$, $g_R$ characterizing the 4d $\mathcal N=4$ SYM theory on the smaller-rank side. 
This leads to (\ref{eq:D3D5-gen-one-point-summary}), (\ref{eq:D3D5-gen-one-point-summary-2}) expressed in terms of field theory data.
The actual one-point functions are obtained from (\ref{eq:mixing-sum});  
we give the first two for $g_L=g_R$ explicitly,
\begin{align}\label{eq:D3D5gen-ex}
	\langle \mathcal O_{J=2}\rangle&=N_{\rm D3}^{\rm R}\left(1-d^2+\kappa ^2\right)\frac{(d-\kappa )^2+3}{6 \sqrt{2}}~,
	\nonumber\\
	\langle \mathcal O_{J=4}\rangle&=
	N_{\rm D3}^{\rm R}\left(d^2-\kappa ^2-1\right) \frac{(d-\kappa ) \left(6 d^3-3 d^2 \kappa -12 d \kappa ^2+5 d+9 \kappa ^3+25 \kappa \right)-15}{60}~.
\end{align}
Alternative expressions can be obtained from the S-dual setups discussed in sec.~\ref{sec:NS5-interface}.
For these one-point functions we did not assume that $\mathfrak N_{\rm D5}$ is small. 
If we add this assumption we can connect to the results of \cite{Komatsu:2020sup} for a single D5-brane, as we explain now.

To make the connection to the defect case we can perform part of the sums in (\ref{eq:D3D5-gen-one-point-2}) using (\ref{eq:sum-hyp-id}) as before. For $\kappa\rightarrow 0$ the hypergeometric function drops out in (\ref{eq:D3D5-gen-one-point-2}) and the result reduces to (\ref{eq:D3D5-defect-one-point-1}).

\textbf{Connecting to \cite{Komatsu:2020sup}:}
In \cite[(3.69)]{Komatsu:2020sup} the one-point functions are given for interfaces with $g_L=g_R$ and a single D5-brane with $k$ D3-branes ending on it, in the `t Hooft large-$N$ limit with $k$ of order $\lambda$. When taking $N\rightarrow\infty$ with fixed $\lambda$, this separates $k$ from $N$.
The S-dual is a single-NS5 interface realizing a bifundamental hypermultiplet.
On the larger-rank side, $\langle \mathcal O_J\rangle = c_J^{(k)}$ with
\begin{align}\label{eq:KW-large-rank}
	c_J^{(k)}&=\left(\kappa+\sqrt{1+\kappa^2}\right)^J\frac{4(-1)^Jg[-\kappa+J\sqrt{1+\kappa^2}]}{2^{J/2}\sqrt{J}(J^2-1)}~,
	&
	\kappa&=\frac{\pi k}{\sqrt{\lambda}}~,
	&
	g&=\frac{\sqrt{\lambda}}{4\pi}~.
\end{align}
The results for the smaller-rank side are obtained up to an overall phase by sending $\kappa\rightarrow -\kappa$.\footnote{As outlined in \cite{Komatsu:2020sup}, for the smaller-rank side the red term in the square brackets in (3.50) there should be omitted, which drops the contribution {\sf extra2} in (3.56). This ends up being equivalent to keeping $\mathsf{extra1}+\mathsf{extra2}$ but sending $J\rightarrow -J$.
Noting that $-\kappa+\sqrt{1+\kappa^2}$ is the inverse of $\kappa+\sqrt{1+\kappa^2}$, this amounts to $\kappa\rightarrow -\kappa$.}
The one-point functions on the smaller-rank side become
\begin{align}\label{eq:KW-small-rank}
	\tilde c_J^{(k)}&=\left(-\kappa+\sqrt{1+\kappa^2}\right)^J\frac{4(-1)^Jg[\kappa+J\sqrt{1+\kappa^2}]}{2^{J/2}\sqrt{J}(J^2-1)}~.
\end{align}

To connect our results for general $N_{\rm D5}$ in (\ref{eq:D3D5-gen-one-point-2}) to these one-point functions we first specialize to equal gauge couplings, still keeping $N_{\rm D5}$ arbitrary, which leads to
\begin{align}\label{eq:K2K3dlambda}
	\frac{4\pi^2 \hat K^2}{\lambda}&=\left(\mathfrak N_{\rm D5}+\sqrt{1+(\mathfrak N_{\rm D5}+\kappa)^2}\right)^2-\kappa^2~,
	&
	\mathfrak N_{\rm D5}&=\frac{\sqrt{\lambda} N_{\rm D5}}{4\pi N_{\rm D3}^{\rm R}}~,
	&
	\kappa&=\frac{\pi k}{N_{\rm D5}\sqrt{\lambda}}~.
\end{align}
This $\kappa$ agrees with the definition in (\ref{eq:KW-large-rank}) when $N_{\rm D5}=1$.
To connect (\ref{eq:D3D5-gen-one-point-2}) to (\ref{eq:KW-small-rank}), we expand (\ref{eq:D3D5-gen-one-point-2}) to linear order in $\mathfrak{N}_{\rm D5}$ while keeping $\kappa$ fixed, and then set $N_{\rm D5}=1$.
We verified that this expansion of (\ref{eq:D3D5-gen-one-point-2}) reproduces (\ref{eq:KW-small-rank}) up to $J=20$.

\bigskip
\textbf{Generalization to multiple D5 groups:}
The discussion can be genearlized straightforwardly to interfaces with $P$ groups of D5-branes,
with $N_{\rm D5}^{(r)}$ D5-branes in the $r^{\rm th}$ group and each D5 in the $r^{\rm th}$ group having $k_r$ D3-branes ending on it. 
This realizes a Nahm pole boundary condition with
\begin{align}
	t_i&=\oplus_{r=1}^P\Big(\oplus_{s=1}^{N_5^{(r)}} t_i^{k_r\times k_r}\Big)~.
\end{align}
The holomorphic functions for the supergravity duals are
\begin{align}
	\mathcal A_1&=\frac{\pi \alpha'}{4}\left(K_0 e^z-K_1 e^{-z}\right)-\frac{i\alpha'}{4}
	\sum_{r=1}^PN_{\rm D5}^{(r)}\ln\tanh\left(\frac{i\pi}{4}-\frac{z-\delta_r}{2}\right)~,
	\nonumber\\
	\mathcal A_2&=\frac{\pi\alpha'}{4}\left(K_2 e^z+K_3e^{-z}\right)~.
\end{align}
The D3-brane numbers and couplings on the left and right half spaces as well as the $k_r$ are determined by the $K_i$ and $\delta_r$, e.g.\ following \cite{Assel:2011xz}.

The field theory still has one set of antisymmetric defect Wilson loops from which the saddle point eigenvalue density can be extracted. Since $\mathcal A_2$ is unchanged compared to (\ref{eq:D3D5-cA12}), the Wilson loops are still described by D5' branes along
\begin{align}
	{\rm D5}': \qquad\Re(z)=\frac{1}{2}\ln\frac{K_3}{K_2}~.
\end{align}
With this information the saddle points can be extracted from (\ref{eq:rho-h12}) and the one-point functions on the smaller-rank side can be determined as before using (\ref{eq:one-point-gen}), (\ref{eq:Mn-def}).

\section{$\mathcal N=4$ SYM BCFTs}\label{sec:BCFT}

The BCFTs we discuss are engineered by D3-branes ending on NS5 and D5 branes. The NS5-branes make manifest that there are boundary degrees of freedom in the form of a 3d $\mathcal N=4$ quiver SCFTs, engineered by Hanany-Witten setups with finite D3-brane segments suspended between the NS5-branes. Depending on the paramaters the D5-branes add flavor hypermultiplets to the 3d quiver or impose partial Nahm pole boundary conditions on the 4d $\mathcal N=4$ SYM fields.

\subsection{D3/NS5 BCFT}\label{sec:D3NS5}

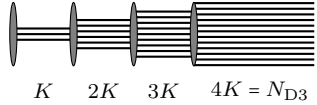
\begin{figure}
		\begin{tikzpicture}[xscale=0.8,yscale=1.5]
			\foreach \i in {-1,0,1} {\draw[thick] (0,{0.05*\i}) -- +(1,0);}
			\foreach \i in {-2.5,-1.5,-0.5,0.5,1.5,2.5} {\draw[thick] (1,{0.05*\i}) -- +(1,0);}
			\foreach \i in {-4,...,4} {\draw[thick] (2,{0.05*\i}) -- +(1,0);}
			\foreach \i in {-5.5,-4.5,...,5.5} {\draw[thick] (3,{0.05*\i}) -- +(2,0);}	
			
			\foreach \i in {0,1,2,3}{\draw[fill=gray] (\i,0) ellipse (1.7pt and 8pt);}
			
			\node at (0.5,-0.5) {\scriptsize $K$};
			\node at (1.5,-0.5) {\scriptsize $2K$};
			\node at (2.5,-0.5) {\scriptsize $3K$};
			\node at (4.1,-0.5) {\scriptsize $4K=N_{\rm D3}$};
		\end{tikzpicture}		
	\caption{Brane constructions for BCFTs, with D3-branes as horizontal lines, NS5-branes as ellipses and D5-branes as vertical lines, for the D3/NS5 BCFT with the quiver in (\ref{eq:D3NS5-quiver}), for $N_5=4$, $K=3$.\label{fig:D3NS5-branes}}
\end{figure}

The first example is engineered by $N_5K$ D3-branes ending on a group of $N_5$ NS5 branes which each have $K$ D3-branes ending on them.
The brane setup is illustrated in fig.~\ref{fig:D3NS5-branes}.\footnote{The D3/NS5 BCFT is S-dual to a limit of the D3/D5 interface of sec.~\ref{sec:D3D5}: Taking $k\rightarrow -N_{\rm D3}^{\rm R}$ in the latter produces a BCFT with Nahm pole boundary condition. The boundary condition breaks part of the gauge symmetry and a global symmetry emerges, leading to new boundary degrees of freedom. These are S-dual to the 3d quiver SCFT.}
The boundary free energy was first computed in \cite{Raamsdonk:2020tin} and aspects of the spectrum were discussed in \cite[sec.~3.2]{Chaney:2024bgx}.
In \cite{He:2024djr} we derived the saddle point dominating the matrix model and confirmed that can be derived independently from the $\rm AdS_4\times S^2\times S^2\times \Sigma$ supergravity dual through the relation in (\ref{eq:rho-h12}).

The field theory engineered by the brane setup is a quiver gauge theory of the form
\begin{align}\label{eq:D3NS5-quiver}
	U(K)-U(2K)-\ldots - U((N_5-1)K)-\widehat{U(N_5K)}
\end{align}
where the node with hat denotes 4d $\mathcal N=4$ SYM and all other nodes are 3d gauge nodes localized on the boundary.
Localization leads to a matrix integral involving one matrix for each gauge node. The saddle point eigenvalue densities were derived independently in field theory and from the holographic duals in \cite{He:2024djr}, with the result in \cite[(50)]{He:2024djr},
\begin{align}\label{eq:rho-tilde-cR}
	\tilde\rho_t(a)&=\Im\left[\mathcal R\left(\frac{2\pi a+i\pi t}{N_5}\right)\right]~,
&
	\mathcal R(v)&=\frac{4N_5}{g_{\rm YM}^2} W_k(-e^{v})
	~,
	&
	k&=\frac{g_{\rm YM}^2K}{4N_5}~.
\end{align}
Here $t=1,\ldots N_5$ denotes the gauge nodes and $a$ denotes the eigenvalues associated with the $t^{\rm th}$ node. This provides the full saddle point. 
For the computation of the one-point function at the north pole of the hemisphere we need the density associated with the 4d $\mathcal N=4$ SYM node, corresponding to $t=N_5$.
That is,
\begin{align}
	\tilde\rho_{\rm 4d}(a)&=\frac{4N_5}{\gym^2}\Im W_k(e^{2\pi |a|/N_5})~,
\end{align}
where the support is $2\pi|a|/N_5<\sqrt{k(k+2)}+\ln\left(k+1+\sqrt{k(k+2)}\right)$ and $W_k$ is a generalized Lambert function \cite{Corless1996OnTL,2014arXiv1408.3999M} defined as in \cite{He:2024djr} by
\begin{align}\label{eq:Wk-def}
	e^{W_k(z)}\frac{W_k(z)+k}{W_k(z)-k}&=z~,
\end{align}
with an appropriate choice of branch and branch cuts, which was also discussed in \cite{He:2024djr}.

The one-point functions can be evaluated using (\ref{eq:one-point-gen}) with (\ref{eq:Mn-def}) or directly using (\ref{eq:one-point-int}).
The moments $M_m$ are, with $a=N_5x/(2\pi)$,
\begin{align}
	M_m&=\int da \,a^{2m}\tilde \rho_{\rm 4d}(a)
	=\left(\frac{N_5}{2\pi}\right)^{2m+1}\frac{4N_5}{\gym^2}\int dx\, x^{2m}\Im W_k(e^x)~.
\end{align}
This integral can be evaluated in the complex plane as in \cite{He:2024djr}. We introduce a $u$ coordinate
\begin{align}\label{eq: u}
	u&=\frac{k-W_k(e^{x})}{k+W_k(e^{x})}~,&
	&\leftrightarrow & 
	x&=k\frac{1-u}{1+u}-\ln (-u)~.
\end{align}
This eliminates the Lambert function in terms of elementary functions.
The remaining integral then reduces to the residue of the integrand at $u=-1$
\begin{align}
	\int dx\, x^{2m}\Im W_k(e^x)
&=\Im \int du~\frac{dx}{du} x^{2m}k\frac{1-u}{1+u}
=\pi k \Re \, \Res_{u=-1}\left[  \frac{dx}{du} x^{2m}\frac{1-u}{1+u}\right].
\end{align}
We thus arrive at
\begin{align}\label{eq:D3NS5-Mn-1}
	M_m&=N_5K\left(\frac{N_5}{2\pi}\right)^{2m}
	\Re \, \Res_{u=-1}\left[\frac{1}{2}\frac{dx}{du} x^{2m}\frac{1-u}{1+u}\right]
	\nonumber\\
	&=\frac{N_5K}{2m+1}\left(\frac{N_5}{2\pi}\right)^{2m}
	 \Res_{w=0}\left[\frac{1}{w^2}\left(\frac{2k}{w}-k-\ln (1-w)\right)^{2m+1}\right],
\end{align}
where integration by parts and $w=1+u$ was used to obtain the second line.
The residues are straightforward to evaluate for arbitrary fixed $m$.
The first, $M_0=N_5K$, reflects the correct normalization of the eigenvalue density.
The auxiliary one-point function in (\ref{eq:one-point-gen}) become, with $N_{\rm D3}=N_5K$,
\begin{align}
	\langle\hat{\mathcal O}_{J=2n}\rangle
	&=
	\frac{N_{\rm D3}}{2^n\sqrt{2n}}
	\left[-\delta_{n,1}
	+
	\sum_{m=0}^{n}\frac{2n(-1)^m\Gamma(m+n)}{\Gamma(2+2m)\Gamma(1-m+n)}
	\Res_{w=0}\left[\frac{\left(\frac{2k}{w}-k-\ln (1-w)\right)^{2m+1}}{w^2k^m}\right]
	\right].
\end{align}
A more compact expression can be obtained directly from (\ref{eq:one-point-int}).
Following the steps above, we find
\begin{align}
	\langle\hat{\mathcal O}_{J}\rangle
	&=
	\frac{N_{\rm D3}}{(-2)^{J/2}\sqrt{J}}\left[\delta_{J,2}+\frac{1}{\pi k}\int dx\, \Im W_k(e^x)\,T_J\left(\frac{x}{2\sqrt{k}}\right)\right]
	\nonumber\\
	&=
	\frac{N_{\rm D3}}{(-2)^{J/2}\sqrt{J}}\left[\delta_{J,2}+\Res_{u=-1}\left[ \frac{dx}{du}\frac{1-u}{1+u}\,T_J\left(\frac{x}{2\sqrt{k}}\right)\right]\right].
	\label{eq:D3NS5-OJ}
\end{align}
Apart from the overall factor of $N_{\rm D3}$ the one-point functions only depend on the combination $k$. 
The residues can be evaluated straightforwardly for fixed $J$.
The actual one-point functions are obtained via (\ref{eq:mixing-sum});
the first examples are
\begin{align}\label{eq:D3NS5-OJ-ex}
	\langle\mathcal O_{J=2}\rangle&=\frac{N_{\rm D3}}{2\sqrt{2}}\left(-\frac{k}{3}-1\right)\,,
	&
	\langle\mathcal O_{J=6}\rangle&=\frac{N_{\rm D3}}{840\sqrt{6}}\left(-15 k^3-161 k^2-525 k-525\right)\,,
	\nonumber\\
	\langle\mathcal O_{J=4}\rangle&=\frac{N_{\rm D3}}{8}\left(\frac{k^2}{5}+\frac{4 k}{3}+2\right)\,,
	&
	\langle\mathcal O_{J=8}\rangle&=\frac{N_{\rm D3}}{10080\sqrt{2}}\left(35 k^4+528 k^3+2744 k^2+5880 k+4410\right)\,.
\end{align}
These one-point functions can be connected to those for the D3/D5 interface in (\ref{eq:D3D5gen-ex}): As discussed in that section, the one-point functions computed there for the smaller-rank side can be continued to the one-point functions on the larger-rank side by taking $k$ negative. The BCFT limit in which the smaller-rank side becomes trivial amounts to $k\rightarrow -N_{\rm D3}$ or $\kappa\rightarrow-1/(4\mathfrak N_{\rm D5})$. In this limit the D3/D5 one-point functions (\ref{eq:D3D5gen-ex}) are related by S-duality to (\ref{eq:D3NS5-OJ-ex}). We verified this for $J=2,4,6,8$.

\subsection{D3/D5/NS5 BCFT}\label{sec:D3D5NS5-BCFT}

The second BCFT example has a group of fundamental hypermultiplets in the 3d part of the D3/NS5 quiver, while keeping all nodes balanced. 
The flavors add a large flavor symmetry group and this class of theories is mapped into itself by S-duality. This also includes the BCFTs that form the basis for the double holography constructions in \cite{Uhlemann:2021nhu,Karch:2022rvr}.

\begin{figure}
	\centering
	\subfigure[][]{\label{fig:D3D5NS5-branes-gen}
	\begin{tikzpicture}[y={(0cm,1cm)}, x={(0.707cm,0.707cm)}, z={(1cm,0cm)}, scale=1.2]
		\draw[gray,fill=gray!100,rotate around={-45:(0,0,1.8)}] (0,0,1.8) ellipse (1.8pt and 3.5pt);
		\draw[gray,fill=gray!100] (0,0,0) circle (1.5pt);
		
		\foreach \i in {-0.05,0,0.05}{ \draw[thick] (0,-1,\i) -- (0,1,\i);}

		\foreach \i in {-0.075,-0.025,0.025,0.075}{ \draw (-1.1,\i,1.8) -- (1.1,\i,1.8);}
		
		\foreach \i in {-0.075,-0.045,-0.015,0.015,0.045,0.075}{ \draw (0,1.4*\i,0) -- (0,1.4*\i,1.8+\i);}
		\foreach \i in  {-0.045,-0.015,0.015,0.045}{ \draw (0,1.4*\i,1.8+\i) -- (0,1.4*\i,4);}
		
		\node at (-0.18,-0.18,3.4) {\footnotesize $N_{\rm D3}$};
		\node at (1.0,0.3,1.8) {\footnotesize $N_{\rm D5}$ D5};
		\node at (0,-1.25) {\footnotesize $N_5$ NS5};
		\node at (0.18,0.18,0.8) {{\footnotesize $N_5R$}};
	\end{tikzpicture}
	}
	\hskip 20mm
		\subfigure[][]{\label{fig:D3D5NS5-branes}
		\begin{tikzpicture}[xscale=0.8,yscale=1.6]
			\foreach \i in {-0.5,0.5} {\draw[thick] (0,{0.07*\i}) -- +(1,0);}
			\foreach \i in {-1.5,...,1.5} {\draw[thick] (1,{0.07*\i}) -- +(1,0);}
			\foreach \i in {-2.5,...,2.5} {\draw[thick] (2,{0.07*\i}) -- +(1,0);}
			\foreach \i in {-2,...,2} {\draw[thick] (3,{0.07*\i}) -- +(1,0);}	
			\foreach \i in {-1.5,...,1.5} {\draw[thick] (4,{0.07*\i}) -- +(2,0);}	
			
			\foreach \i in {0,...,4}{\draw[fill=gray] (\i,0) ellipse (1.7pt and 8pt);}
			
			\foreach \i in {-0.15,0,0.15} {\draw[thick] (2.5+\i,-0.7) -- +(0,1.4);}
			
			\node at (0.5,-0.5) {\scriptsize $R$};
			\node at (1.5,-0.5) {\scriptsize $2R$};
			\node at (5.5,-0.5) {\scriptsize $N_{\rm D3}$};
			
			\node at (2.5,-0.85) {\scriptsize $N_{\rm D5}$};
		\end{tikzpicture}	
	}
	\caption{Left: the D3/D5/NS5 BCFT engineered by $N_{\rm D3}$ semi-infinite D3-branes ending on a combination of $N_{\rm D5}$ D5-branes and $N_5$ NS5-branes, as described in the text. Right after Hanany-Witten transitions to make the quiver in (\ref{eq:D5NS5K-quiver}) manifest, for $N_5=5$, $N_{\rm D5}=3$, $R=2$, $S=-2$.
		\label{fig:brane-D5NS5-D3}}
\end{figure}
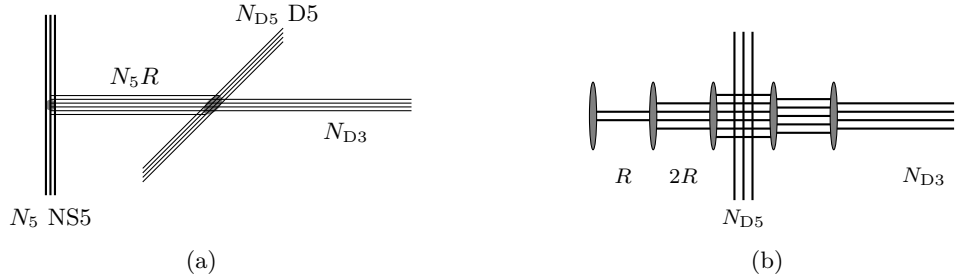

The brane construction, illustrated in fig.~\ref{fig:D3D5NS5-branes-gen}, involves an additional group of D5 branes and we refer to this as D3/D5/NS5 BCFT.
It has $N_{\rm D3}$ semi-infinite D3-branes ending on $N_5$ NS5 branes and $N_{\rm D5}$ D5-branes, in such a way that each NS5-brane has $R$ D3-branes ending on it from the right and each D5-brane has $S$ D3-branes ending on it from the right.
$S$ can be negative, then $|S|$ D3-branes end on the D5-brane from the left.
The number of D3-branes is 
\begin{align}
N_{\rm D3}&=RN_5+SN_{\rm D5}~.
\end{align}

For $S>0$ the D5-branes impose partial Nahm pole boundary conditions while the remaining 4d $\mathcal N=4$ SYM fields couple to a 3d SCFT with no flavors. For $S<0$ the D5-branes can be moved by Hanany-Witten transitions to a location between NS5-branes so that they have no net D3-branes ending on them. This leads to the form in fig.~\ref{fig:D3D5NS5-branes}.
The entire 4d $\mathcal N=4$ SYM gauge group couples to a 3d SCFT, which now contains fundamental hypermultiplets engineered by the D5-branes, by gauging a global symmetry. This transition was discussed in \cite{Karch:2022rvr}, \cite[sec.~4.4]{Chaney:2024bgx}. 

Here we focus on $S<0$. The field theory is then described by the balanced quiver
\begin{align}\label{eq:D5NS5K-quiver}
	U(R)-U(2R)-\ldots &- U(TR) - U(TR-Q)-\ldots - U(N_{\rm D3}+Q) - \widehat{U(N_{\rm D3})}
	\nonumber\\
	&\ \ \ \ \ \ \,\vert\\
	\nonumber & \ \ \ [N_{\rm D5}]	
\end{align}
where
\begin{align}
	T&=N_{5}+S~,
	&
	Q&=N_{\rm D5}-R~.
\end{align}
Along the first ellipsis in (\ref{eq:D5NS5K-quiver}) the ranks increase in steps of $R$,
along the second they decrease in steps of $Q$.
The total number of gauge nodes is $N_5$ and all 3d nodes are balanced.

The field theory has 5 parameters: $N_{\rm D3}$ and $\gym^2$ characterize 4d $\mathcal N=4$ SYM, the 5-brane numbers $N_5$ and $N_{\rm D5}$ specify the global symmetries $U(N_5)$ and $U(N_{\rm D5})$, and there is one additional parameter, which can, for example, be taken as the initial slope of the 3d ranks, $R$, or as the position of the fundamental fields within the 3d quiver $T$.

The matrix model was discussed in \cite{He:2024djr}. 
The saddle point eigenvalue densities are, with $t$ labeling the gauge nodes as before \cite[(9)]{He:2024djr},
\begin{align}\label{eq:rhot-D3D5NS5-sum}
	\tilde\rho_t(a)&=\Im\left[\mathcal R\left(\frac{2\pi a+i\pi t}{N_5}\right)\right]~,
	\nonumber\\
	\mathcal R(v)&=
	\frac{4N_5}{\gym^2}
	W_k(-e^{v})
	-\frac{i}{\pi} N_{\rm D5}\ln\left(\frac{W_k(-e^{v})-W_k(-e^{i\pi T/N_5})}{W_k(-e^{v})+W_k(- e^{i\pi T/N_5})}\right)~,
\end{align}
where $k$ is determined in terms of field theory data by imposing the correct normalization for the densities.
The eigenvalue density for the 4d gauge node with $t=N_5$ becomes, with $a=N_5x/(2\pi)$,
\begin{align}
	\tilde\rho_{\rm 4d}\left(\frac{N_5x}{2\pi}\right)&=\Im\left[
	\frac{4N_5}{\gym^2}
	W_k(e^{x})
	-\frac{i}{\pi} N_{\rm D5}\ln\left(\frac{W_k(e^{x})-W_k\left(-e^{i\pi T/N_5}\right)}{W_k(e^{x})+W_k\left(-e^{i\pi T/N_5}\right)}\right)
	\right]\,.
\end{align}
The support at the 4d node is $|x|<\sqrt{k(k+2)}+\ln\left(k+1+\sqrt{k(k+2)}\right)$.\footnote{Physically, $k$ thus sets the expectation value of a circular Wilson loop on the boundary of the BCFT on a hemisphere, in the fundamental representation of the 4d node, as
\begin{align}
	\langle W_f\rangle&=N_5\left(\sqrt{k(k+2)}+\ln\left(k+1+\sqrt{k(k+2)}\right)\right)\,.
\end{align}}

The remaining parameter $k$ can be fixed through the normalization of $\tilde\rho$, and we can equivalently determine it from \cite[(8)]{He:2024djr}, which provides a reparametrization of $R$ and $S$ in terms of $k$ and $\delta$ (which appear naturally in the supergravity solution) as
\begin{align}\label{eq:RS-sum}
	R&=\frac{4N_5}{\gym^2}k+\frac{2}{\pi}N_{\rm D5} \cot ^{-1}\!e^{\delta }~, & 
	S&=\frac{N_5}{\pi}\left(ke^\delta-2\cot^{-1}\!e^\delta\right) ~.
\end{align}
It will be convenient to express $R$ and $S$ in terms of $T$, i.e.\ the position of the flavors in the 3d quiver. The equations can then be expressed as 
\begin{align}\label{eq:T-k-delta}
	\frac{\pi T}{N_5}&=k e^\delta+2\tan^{-1}e^\delta~,
	&
	e^\delta k&=\frac{N_{\rm D3}}{N_5N_{\rm D5}}\left(1-\frac{4kN_5^2}{\lambda}\right)~.
\end{align}
A useful relation which also appeared in \cite[(49)]{He:2024djr} is
\begin{align}
	W_k\left(-e^{i\pi T/N_5}\right)=ike^\delta~,
\end{align}
where we note that the Lambert function of a phase is imaginary.
Solving the second equation in (\ref{eq:T-k-delta}) for $\delta$ leaves a transcendental equation for $k$ in terms of field theory data as
\begin{align}
	\frac{\pi T}{N_5}&=\frac{N_{\rm D3}}{N_5N_{\rm D5}}\left(1-\frac{4kN_5^2}{\lambda}\right)+2\tan^{-1}\left(\frac{N_{\rm D3}}{N_5N_{\rm D5}}\left(\frac{1}{k}-\frac{4N_5^2}{\lambda}\right)\right)~.
\end{align}
The left hand side is valued in $(0,\pi)$ and the right side is monotonic in $k$ as both terms increase with decreasing $k$, so the solution is unique. For $T/N_5\rightarrow 0$ we have $k\rightarrow \lambda/(4N_5^2)$.

With these ingredients the moments and one-point functions can be determined following the same strategy as for the D3/NS5 BCFT.
The moments become
\begin{align}
	M_m=\int da \,a^{2m}\tilde \rho_{\rm 4d}(a)
	&=\left(\frac{N_5}{2\pi}\right)^{2m+1}\int dx\, x^{2m}\Im\left[\frac{4N_5}{\gym^2} W_k(e^x)
	-\frac{i}{\pi} N_{\rm D5}\ln\left(\frac{W_k(e^{x})-ike^\delta}{W_k(e^{x})+ike^\delta}\right)\right]
	\nonumber\\
	&=\left(\frac{N_5}{2\pi}\right)^{2m+1} \pi \Res_{u=-1}\left[  \frac{dx}{du} x^{2m}\left(
	\frac{4 N_5k}{\gym^2} \frac{1-u}{1+u}
	-\frac{i}{\pi}N_{\rm D5}\ln\left(\frac{\frac{1-u}{1+u}-ie^\delta}{\frac{1-u}{1+u}+ie^\delta}\right)\right)\right],
	\label{eq:Mn-D3D5NS5}
\end{align}
where the coordinate transformation in (\ref{eq: u}) to eliminate the Lambert function in terms of elementary functions in the second line.
The expression in (\ref{eq:Mn-D3D5NS5}) is straightforward to evaluate for fixed $m$.
The first two moments are
\begin{align}
	M_0&=k N_5\left(\frac{N_{\rm D5}e^\delta}{\pi }+\frac{4N_5}{g^2}\right)
	=N_{\rm D3}~,
	&
	M_1&=k^2N_5^3\left(\frac{(k+6)N_5}{3 \pi ^2 g^2}+\frac{N_{\rm D5}(6e^\delta-e^{3\delta} k)}{12 \pi ^3}\right)~,
\end{align}
where $M_0$ gives the normalization as expected with $k$ determined as above. 
The auxiliary one-point functions (\ref{eq:one-point-int}) become
\begin{align}
	\langle\hat{\mathcal O}_J\rangle&=\frac{1}{(-2)^{J/2}\sqrt{J}}\left[N_{\rm D3}\delta_{J,2}+\frac{N_5}{\pi}\int dx\, \tilde\rho_{4d}\left(\frac{N_5x}{2\pi}\right)\,T_J\left(\frac{N_5 x}{\sqrt{\lambda}}\right)\right],
\end{align}
where the remaining integral can be evaluated as for the moments before,
\begin{align}
	N_5\int dx\, &\tilde\rho_{4d}\left(\frac{N_5x}{2\pi}\right)\,T_J\left(\frac{N_5 x}{\sqrt{\lambda}}\right)
	\nonumber\\
	&=N_{\rm D3}\Im\int dx\left[\frac{4N_5^2}{\lambda}W_k(e^x)-\frac{i}{\pi}\frac{N_{\rm D5}N_5}{N_{\rm D3}}
	\ln\left(\frac{W_k(e^x)-ik e^\delta}{W_k(e^x)+ik e^\delta}\right)
	\right]T_J\left(\frac{N_5 x}{\sqrt{\lambda}}\right)\,.
\end{align}
This leads to 
\begin{align}\label{eq:one-point-D3D5NS5-0}
	\langle\hat{\mathcal O}_J\rangle&=
	\frac{N_{\rm D3}}{(-2)^{J/2}\sqrt{J}}\left[\delta_{J,2}
	+
	\Res_{u=-1}\left[\frac{dx}{du}
	\left(
	\frac{4N_5^2k}{\lambda}\frac{1-u}{1+u}
	-\frac{i}{\pi}\frac{N_{\rm D5}N_5}{N_{\rm D3}}
	\ln\left(\frac{\frac{1-u}{1+u}-ie^\delta}{\frac{1-u}{1+u}+ie^\delta}\right)
	\right)
	T_J\left(\frac{N_5 x}{\sqrt{\lambda}}\right)
	\right]
	\right].
\end{align}
For the 5-brane numbers and `t Hooft couplings two combinations that appear naturally are 
\begin{align}\label{eq:N5ND5-def}
	\mathfrak N_5&=\frac{N_5}{\sqrt{\lambda}}~, & 
	\mathfrak{N_{\rm D5}}&=\frac{N_{\rm D5}\sqrt{\lambda}}{4\pi N_{\rm D3}}~.
\end{align}
They are exchanged by S-duality.
Then
\begin{align}\label{eq:one-point-D3D5NS5}
	\langle\hat{\mathcal O}_J\rangle&=
	\frac{N_{\rm D3}}{(-2)^{J/2}\sqrt{J}}\left[\delta_{J,2}
	+
	4\mathfrak{N}_5
	\Res_{u=-1}\left[\frac{dx}{du}
	\left(
	\mathfrak N_5k\frac{1-u}{1+u}
	-i \mathfrak{N}_{\rm D5}
	\ln\left(\frac{\frac{1-u}{1+u}-ie^\delta}{\frac{1-u}{1+u}+ie^\delta}\right)
	\right)
	T_J\left(\mathfrak{N}_5x\right)
	\right]
	\right].
\end{align}
where we recall from (\ref{eq:T-k-delta})
\begin{align}
	ke^\delta&=\frac{1-4 k\mathfrak{N}_5^2}{4\mathfrak{N}_5\mathfrak{N}_{\rm D5}}~,
	&
	\frac{\pi T}{\sqrt{\lambda}\mathfrak{N}_5}&=e^{\delta } k+
	2 \tan ^{-1}e^{\delta}~.
\end{align}
The $\langle \hat{\mathcal O}_J\rangle$ depend on $N_{\rm D3}$ through the overall factor and are otherwise expressed in terms of the combinations $k$, $\mathfrak{N}_5$, $\mathfrak{N}_{\rm D5}$ and $T/\sqrt{\lambda}$.
The first examples for the actual one-point functions (\ref{eq:mixing-sum}) are
\begin{align}\label{eq:D3D5NS5-ex}
	\langle O_{J=2}\rangle&=N_{\rm D3}\frac{3-8 k \mathfrak{N}_5^2 \left(2 k^2 \mathfrak{N}_5 \left(e^{3 \delta } \mathfrak{N}_{\rm D5}+\mathfrak{N}_5\right)+3\right)}{6 \sqrt{2}}~,
	\nonumber\\
	\langle O_{J=4}\rangle&=\frac{16 k \mathfrak{N}_5^2 \left(6 k^3 (k+10) \mathfrak{N}_5^4-5 (k-12) k \mathfrak{N}_5^2-15\right)+15}{60}N_{\rm D3}
	\nonumber\\&\hphantom{=}\
	-\frac{24 k \mathfrak{N}_5^2 \left(e^{2 \delta } k-10\right)+40}{15} e^{3 \delta } k^3\mathfrak{N}_{\rm D5} \mathfrak{N}_5^3
	N_{\rm D3}~.
\end{align}
In the `t Hooft limit for the ambient 4d $\mathcal N=4$ SYM theory, $N\rightarrow\infty$ at fixed $\lambda$,
keeping $\mathfrak{N}_5$ and $\mathfrak N_{\rm D5}$ finite amounts to keeping $N_5$ fixed and $N_{\rm D5}\sim N_{\rm D3}$ when taking $N_{\rm D3}\rightarrow\infty$, and then $N_5\sim\sqrt\lambda$ and $N_{\rm D5}/N_{\rm D3}\sim 1/\sqrt{\lambda}$ in the strong-coupling limit of large $\lambda$.
It may be interesting to also investigate the `very strong coupling' limit of taking $N\rightarrow\infty$ at fixed $\gym$ \cite{Azeyanagi:2013fla,Binder:2019jwn}.

\textbf{Special case $\delta=0$:} 
Imposing $\delta=0$ leads to the family of theories discussed in \cite[sec.~4]{Karch:2022rvr}. This adds a relation between the 4d gauge coupling and the 5-brane numbers. This family is the S-duality orbit of the theory used in \cite{Uhlemann:2021nhu} for black hole studies (the entanglement entropy computations of \cite{Uhlemann:2021nhu} are insensitive to $SL(2,\ZZ)$).
For $\delta=0$,
\begin{align}
	k&=\frac{1}{4\mathfrak{N}_5(\mathfrak{N}_5+\mathfrak{N}_{\rm D5})}~,
\end{align}
and the one-point functions (\ref{eq:mixing-sum}) only depend on $\mathfrak{N}_5$ and $\mathfrak{N}_{\rm D5}$.
The first 3 are, with $\mathfrak{N}_\pm=\mathfrak{N}_{\rm D5}\pm \mathfrak{N}_5$,
\begin{align}\label{eq:D3D5NS5-delta0-ex}
	\langle O_{J=2}\rangle&=N_{\rm D3}\mathfrak{N}_-\frac{12 \mathfrak{N}_+^2+1}{24 \sqrt{2} \mathfrak{N}_+^3}~,
	\qquad\qquad
	\langle O_{J=4}\rangle=N_{\rm D3}\frac{60 \mathfrak{N}_-^2 \left(8 \mathfrak{N}_+^2+1\right)+20 \mathfrak{N}_+^2+3}{1920
		\mathfrak{N}_+^4}~,
	\nonumber\\
	\langle O_{J=6}\rangle&=N_{\rm D3}\mathfrak{N}_-\frac{28  \mathfrak{N}_+^2 \left(60 \mathfrak{N}_-^2 \left(20  \mathfrak{N}_+^2+3\right)+120  \mathfrak{N}_+^2+23\right)+15}{53760
		\sqrt{6}  \mathfrak{N}_+^7}~,
	\nonumber\\
	\langle O_{J=8}\rangle&=N_{\rm D3}\frac{7 \left(216\mathfrak{N}_-^2+5\right)+2240 \left(84 \left(6 \mathfrak{N}_-^4+\mathfrak{N}_-^2\right)+1\right) \mathfrak{N}_+^4+24
		\left(7840 \mathfrak{N}_-^4+1736 \mathfrak{N}_-^2+25\right) \mathfrak{N}_+^2}{2580480 \sqrt{2} \mathfrak{N}_+^8}.
\end{align}
The D3/NS5 BCFT results in (\ref{eq:D3NS5-OJ}), (\ref{eq:D3NS5-OJ-ex}) are recovered for $\mathfrak{N}_-\rightarrow -\mathfrak{N}_+$. One recovers the expression for $k$ in the D3/NS5 BCFT as $\mathfrak{N}_+= 1/(2\sqrt{k})$ and (\ref{eq:one-point-D3D5NS5}) reduces to (\ref{eq:D3NS5-OJ}).

S-duality acts on the BCFT by exchanging $\mathfrak{N}_{\rm D5}$ and $\mathfrak{N}_5$, which amounts to taking $\mathfrak{N}_-\rightarrow -\mathfrak{N}_-$. 
Under this operation the one-point functions in (\ref{eq:D3D5NS5-delta0-ex}) transform as $\langle \mathcal O_J\rangle \rightarrow (-1)^{J/2}\langle \mathcal O_J\rangle$.
The sign can be understood from the fact that swapping $\mathfrak{N}_{\rm D5}\leftrightarrow\mathfrak{N}_5$ takes the BCFT to its S-dual, but keeps the operator unchanged, while S-duality also acts on the operators, and we find the expected behavior under S-duality.

The discussion above can be generalized straightforwardly to general balanced quivers as boundary degrees of freedom. The saddle points were discussed explicitly in \cite{He:2024djr} and the one-point functions are special case of the discussion in sec.~\ref{sec:NS5-interface}.

\section{D3/D5/NS5 Interface}\label{sec:NS5-interface}

In this section we study interfaces hosting 3d quiver SCFTs, engineered by D3-branes intersecting NS5 and D5 branes.
These are generalizations of the BCFTs studied in sec.~\ref{sec:BCFT}.
We discuss the matrix models and derive the saddle point eigenvalue densities from the supergravity duals.
When there are no D5-branes, these SCFTs are the S-dual to the setups of sec.~\ref{sec:D3D5}.


\subsection{Matrix models}

We start with the matrix models for the general class of 4d $\mathcal N=4$ SYM ICFTs with 3d $\mathcal N=4$ interface SCFTs. 
The matrix models are constructed by combining matrix models for 4d $\mathcal N=4$ SYM on a hemisphere \cite{Gava:2016oep,KumarGupta:2019nay} with matrix models for 3d $\mathcal N=4$ gauge theories \cite{Kapustin:2009kz,Kapustin:2010xq,Benvenuti:2011ga} using gluing formulas \cite{Dedushenko:2018aox,Dedushenko:2018tgx}.
We will be brief since much of the general discussion follows \cite{He:2024djr} where the saddle point equations were derived and solved for BCFTs.
We consider quivers of the form\footnote{We focus on cases without (partial) Nahm pole boundary conditions, where all D5-branes realize flavors in the 3d part of the quiver. This can be generalized (see the comments in sec.~\ref{sec:D3D5NS5-BCFT}).}
\begin{align}\label{eq:quiver-gen}
	&\widehat{U(N_1)}-U(N_2)-\ldots - U(N_{L-1})-\widehat{U(N_L)}~.
	\nonumber\\
	&\hskip 5mm  \hskip 15mm \vert \hskip 25mm\vert
	\\
	&\hskip 8mm \hskip 9mm [k_2] \hskip 5mm \ldots \hskip 7mm [k_{L-1}]
	\nonumber
\end{align}
The hatted nodes denote two 4d $\mathcal N=4$ SYMs on half spaces with gauge groups $U(N_1)$ and $U(N_L)$, all others are 3d nodes. The dashes denote 3d hypermultiplets in the bifundamental representation of the nodes they connect. $[k_t]$ denotes the flavor symmetry of $k_t$ hypermultiplets in the fundamental representation of a gauge node.
The matrix model is
\begin{align}\label{eq:matrix-model-gen}
	\mathcal Z=\,& \frac{1}{N_{1} ! \ldots N_{L} !} \int\left(\prod_{t=1}^{L} \prod_{i=1}^{N_t} d a_{t,i}\right) 
	e^{-\frac{4 \pi^2}{g_{ R}^2} \sum_{i=1}^{N_L} a_{L, i}^2}
	e^{-\frac{4 \pi^2}{g_{L}^2} \sum_{i=1}^{N_1} a_{1, i}^2}\nonumber\\
	&\prod_{i<j}^{N_L}\left(a_{L, i}-a_{L, j}\right) 2\sinh\left(\pi\left(a_{L, i}-a_{L, j}\right) \right)
	\prod_{i<j}^{N_1}\left(a_{1, i}-a_{1, j}\right) 2\sinh\left(\pi\left(a_{1, i}-a_{1, j}\right) \right)
	\nonumber\\
	& \prod_{t=1}^{L-1} \prod_{i<j}^{N_t} 4\sinh^2\left(\pi\left(a_{t,i}-a_{t,j}\right) \right)
	\prod_{t=1}^{L-1} \prod_{i=1}^{N_t} \prod_{j=1}^{N_{t+1}} \frac{1}{2\cosh\left(\pi\left(a_{t,i}-a_{t+1,j}\right)\right)}
	\prod_{t=2}^{L-1} \prod_{i=1}^{N_t} \frac{1}{2^{k_t}\cosh^{k_t}(\pi a_{t,i})}~,
\end{align}
where the first and second lines arise from the two 4d half-space SYM theories, and the last line represents the 3d vector multiplets and fundamental and bifundamental hypermultiplets.
In the long-quiver large-$N$ limit the field theory is described by continuous data
\begin{align}
	z&=\frac{t}{L}~, 
	&
	\lbrace N_t\rbrace &\rightarrow N(z)~,
	&
	k(z)=\sum_{t=1}^L \frac{k_t}{L} \delta\left(z-z_t\right)~,
\end{align}
such that $z\in (0,1)$ is an effectively continuous coordinate along the quiver, the ranks are encoded in an effectively continuous function $N(z)$.
The saddle point dominating the matrix model is encoded in a family of eigenvalue densities which is captured by a function of two real variables,
\begin{align}\label{eq: continuelimit}
	\tilde\rho(z,a)=\tilde\rho_t(a)&=\sum_{i=1}^{N_t} \delta\left(a-a_{t,i}\right)~.
\end{align}
The saddle point equations reduce to a 2d electrostatics problem.
To describe the 2d problem we introduce $\vrho(z,x)=N_5 \trho(z, N_5 x)$. For $z\in (0,1)$, $\varrho$ satisfies a Laplace equation with sources given by the flavors and supplemented by appropriate normalization conditions,
\begin{equation}\label{eq:saddle-bulk}
	\partial^2_ x \vrho(z, x)+4 \partial ^2_z \vrho(z, x)+4L^2 \delta(x)k(z)=0~.
\end{equation}
At the two boundaries $z_b=0,1$ we have two singular integral equations, which are essentially two copies of the condition for the BCFT studied in \cite{He:2024djr}
\be\label{eq: z=1}
(-1)^{z_b}\pi\int_{-x}^x  dy~ \p_z\vrho(z,y)\big\vert_{z=z_b}+\frac{8\pi^2L^2}{\gym^2}x-\int dy\frac{\vrho(z_b,y)}{x-y}=0~.
\ee
The form of solutions depends on whether the 3d part of the theory is balanced or unbalanced; in either case it can be derived from the supergravity duals via \eqref{eq:rho-h12}.

\subsection{Saddle points}\label{sec:NS5-interface-saddle}
We now introduce the gravity duals of the interfaces described above.
For simplicity we focus on general balanced 3d quivers, which are engineered by multiple groups of D5 branes and a single group of NS5 branes. Including unbalanced nodes amounts to allowing multiple NS5-brane groups, and the generalization is straightforward.
The holomorphic functions for balanced 3d quivers are\footnote{We used $z$ to label the nodes in the quiver diagrams before and here as complex coordinate on $\Sigma$ in the supergravity solutions. We hope the meaning will be clear from context.}
\begin{align}\label{eq: a1-a2-inter-NS5-gen}
	\mathcal A_1&=\frac{\pi \alpha'}{4}\left(K_0 e^z-K_1 e^{-z}\right)-\frac{i\alpha'}{4}
	\sum_{d=1}^P N_{\rm D5}^{(d)}\ln\tanh\left(\frac{i\pi}{4}-\frac{z-\delta_d}{2}\right)~,
	\nonumber\\
	\mathcal A_2&=\frac{\pi\alpha'}{4}\left(K_2 e^z+K_3e^{-z}\right)-\frac{\alpha'}{4}N_5 \ln \tanh(\frac{z}{2})         ~.
\end{align}
Asymptotic $\rm AdS_5\times S^5$ regions emerge at $\Re(z)\rightarrow\pm\infty$. 
The asymptotic dilaton and numbers of semi-infinite D3-branes can be obtained from the behavior of the supergravity fields in these regions. This yields
\begin{align}
	e^{2\phi_L}&=\frac{K_3}{K_1}= \frac{g_L^2}{4\pi}~,
	&
	N_{\rm D3}^{\rm R}&=\frac{\pi}{2} (K_0K_3+K_1K_2)+ K_0 N_5+  K_2\sum_de^{\delta_d} N_{\rm D5}^{(d)}~,
	\nonumber\\
	e^{2\phi_R}&=\frac{K_2}{K_0}= \frac{g_R^2}{4\pi}~,
	&
	k\equiv~ N_{\rm D3}^{\rm L}-N_{\rm D3}^{\rm R}&=\sum_de^{\delta_d}N_{\rm D5}^{(d)}(e^{-\delta_d}K_3-e^{\delta_d}K_2)+N_5(K_1-K_0)~.
\end{align}
The details of the 3d quiver SCFT are encoded in the numbers of D5 and NS5-brane groups and the locations of the corresponding sources on $\Sigma$. 
We will discuss this explicitly for examples below.

The saddle point eigenvalue densities can be extracted from the general relation (\ref{eq:rho-h12}) without spelling out the matrix models explicitly.
We now analyze their properties.
The complex combination $v=(2\pi a+i\pi t)/N_5$ of the real matrix model variables $a$ and $t$ is defined from $\mathcal A_2$,\begin{align}\label{eq: z-to-v-NS5-inter}
	v&=k_2 e^z+k_3 e^{-z}-\ln \tanh\left(\frac{z}{2}\right)+i\pi~,
	&
	k_{i}&\equiv \frac{\pi K_{i}}{N_5}~.
\end{align}
The inverse of this relation defines a function $W_{k_2,k_3}(-e^v)$, similarly to the generalized Lambert function in (\ref{eq:Wk-def}), that can be used to give explicit expressions for the eigenvalue densities,
\begin{align}
	\frac{w+1}{w-1}e^{k_2 w+\frac{k_3}{w}}\Big\vert_{w=W_{k_2,k_3}(-e^v)}&=-e^v~,
	&
	e^z&=W_{k_2,k_3}(-e^v)~.
\end{align}
Compared to the Lambert functions in (\ref{eq:Wk-def}), this contains an extra exponential.
The two harmonic functions can now be expressed as
\begin{align}
	h_1&=-\frac{i\alpha'}{4}N_5\left(e^{-2\phi_R}k_2 W_{k_2,k_3}(-e^v)-\frac{e^{-2\phi_L}k_3}{ W_{k_2,k_3}(-e^v)}\right)
	-\frac{\alpha'}{4}\sum_{p=1}^P N_{\rm D5}^{(p)}\ln\left(\frac{W_{k_2,k_3}(-e^v)-i e^{\delta_p}}{W_{k_2,k_3}(-e^v)+ie^{\delta_p}}\right)+{\rm c.c.}
	\nonumber\\
	h_2 &=\frac{\alpha'}{4}N_5 v+{\rm c.c.}
\end{align}
and the eigenvalue density is, with $v=(2 a+i t)\pi/N_5$,
\bea\label{eq: saddle-D3/D5/NS5-gen}
\trho_t(a)&=\Im\left [ K_0 W_{k2,k3}(-e^v)-\frac{K_1}{W_{k2,k3}(-e^v)}
-\frac{i}{\pi}\sum_{p=1}^P N_{\rm D5}^{(p)}\ln\left(\frac{W_{k_2,k_3}(-e^v)-i e^{\delta_p}}{W_{k_2,k_3}(-e^v)+ie^{\delta_p}}\right)\right]~.
\eea

\begin{figure}
	\centering
	\subfigure[][]{\label{fig:WL-D5-NS5-K-a}
		\begin{tikzpicture}
			\node at (0,0) {\includegraphics[width=0.33\linewidth]{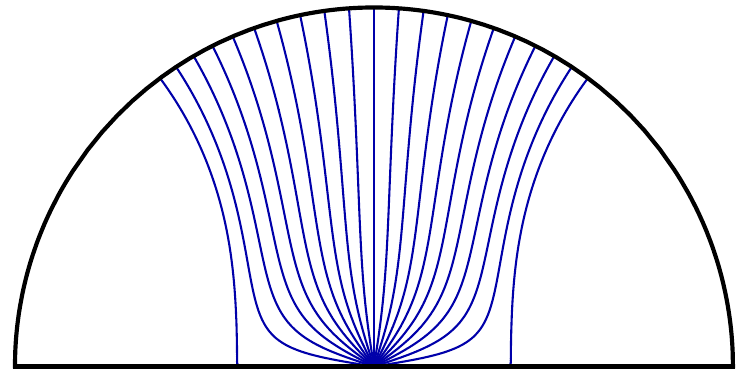}};
			\node at (0.7,0.7) {$\mathbf{\Sigma}$};
			\draw[very thick] (0,-1.2) -- (0,-1.45) node [anchor=north]{\footnotesize NS5};
			\draw[very thick] (0,1.2) -- (0,1.45) node [anchor=south]{\footnotesize D5};
			\draw[very thick] (1,1) -- (1.1,1.25) node [anchor=south]{\footnotesize D5};
			\draw[very thick] (-1,1) -- (-1.1,1.25) node [anchor=south]{\footnotesize D5};
			
			\draw[very thick] (-2.7,-1.4) -- (-2.5,-1.2);
			\node at (-2.75,-1.6) {\footnotesize D3};
			
			\draw[very thick] (2.7,-1.4) -- (2.5,-1.2);
			\node at (2.75,-1.6) {\footnotesize D3};
			
			\node at (-0.95,-1.5) {\footnotesize F1};
			\node at (0.95,-1.5) {\footnotesize F1};
	\end{tikzpicture}	}
	\caption{$\Sigma$ as half disc in the $u$ coordinate with Wilson loop D5$^\prime$ embeddings as blue curves.
		Those emerging from the NS5 source describe Wilson loops associated with 3d gauge nodes ($0<z<1$).
		The embeddings emerging from the points marked F1 describe Wilson loops associated with the two 4d gauge nodes ($z=0$ and $z=1$).
        The eigenvalue densities $\vrho(z,x)$ can be obtained from the corresponding D5$^\prime$ branes. The points marked F1 are solutions to $\frac{dv}{du}=0$; they host fundamental strings describing Wilson loops (for details we refer to \cite{Coccia:2021lpp}). These are also branch points in the map between the $u$ and $v$ coordinates. \label{fig:ICFT temp}
		On each side the line extending between F1 and the D3, and the line between F1 and the NS5, are mapped to the same branch cut in the $v$ plane.}
\end{figure}
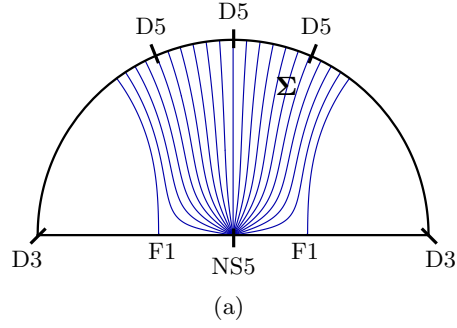

We now discuss the branch cuts and analytic structure.
To this end we introduce a $u$ coordinate as follows,
\begin{align}\label{eq:u-z-v-NS5-inter}
	e^z &=\frac{1-u}{1+u}~,
	&
	v &=k_2 \frac{1-u}{1+u}+k_3 \frac{1+u}{1-u}-\ln u~.
\end{align}
This maps the strip $\Sigma(z,\bar{z})$ to the lower half unit disk.
The result is shown in fig.~\ref{fig:ICFT temp} along with a sample of D5$^\prime$ branes which describe antisymmetric Wilson loops embedded along $h_2^D=\rm const$, as in \cite{Coccia:2021lpp, He:2024djr}.
The map $u\rightarrow v$ is locally a conformal map, except at points where the derivative vanishes, which give the branch points
\bea\label{eq:deri-NS5-gen}
\frac{dv}{du}=-\frac{2 k_2}{(u+1)^2}+\frac{2 k_3}{(u-1)^2}-\frac{1}{u}=0~.
\eea
This is a quartic equation for $u$.
The four roots come in by pairs, denoted by subscript $0$ and $1$. 
In each pair the two roots are the inverse of each other, so
\begin{align}\label{eq: root-deri}
	u_{0+}u_{0-}=u_{1+}u_{1-}&=1~,&
	u_0>0~,\quad u_1<0~.
\end{align}
Here we adapt the convention that $|u_+|<1$ and $|u_-|>1$.
The solutions can be expressed concisely in terms of a `Zukhovsky variable' $y$,
\begin{align}
	y&=u+\frac{1}{u}~,
	&
	\deri{v}{u}&=\frac{y^2+k_-y-1 - k_+}{(1 - y^2)(y + \sqrt{-1 + y^2})}=0~,
	&
	k_\pm&=k_2 \pm k_3~.
\end{align}
The solutions are, with $b=0,1$
\begin{align}
	u_{b,\pm}&=\frac{1}{2}(y_b\pm\sqrt{y_b^2-4})~,
	&
	y_b &= \frac{1}{2} \left(-k_- +(-1)^{b+1} \sqrt{4 + 4k_+ + k_-^2}\right)~.
\end{align}
The branch points of $W_{k_2,k_3}(e^{-v})$ are also given by the four roots derived above.\footnote{The map $v\rightarrow W_{k_2,k_3}(e^{-v})$ can be decomposed into $v\rightarrow u$ and $u\rightarrow W_{k_2,k_3}(e^{-v})\equiv\frac
	{1-u}{1+u}$. With the latter conformal except at $u=\pm 1$, the relevant branch points are those of the former.}
The branch cuts extend from $v(u_{b,\pm})$ to $\pm\infty$ along $\Im{v}=b\pi$.
Between the four cuts is the strip $\mathbb{R}\times [0,\pi]$ in the $v$ plane where $W_{k_2,k_3}(e^{-v})$ and $\vrho(v)$ are defined.
The relevant branch of $W_{k_2,k_3}$ can be inferred from the supergravity construction and is distinguished by $W_{k_2,k_3}(e^{-\overline v})=\overline{W_{k_2,k_3}(e^{- v})}$ in this domain.

We now discuss the eigenvalue densities and derive general expressions for the one-point functions. 
The support of $\vrho(v)$ is also determined by the branch points.
From \eqref{eq: saddle-D3/D5/NS5-gen} one can see that the eigenvalue densities are only non-vanishing when $u$ has a non-vanishing imaginary part.
Therefore, in the interior $0<\Im{v}<\pi$, they have non-compact support. This is reflected in fig.~\ref{fig:ICFT temp} in which the corresponding D5$^\prime$ branes emerge from the origin where $v(0)$ is infinity. 
For the boundary nodes it has compact support $(v(u_{b-}),v(u_{b+}))$, with the end points coinciding with the branch points.

Integrals involving $\vrho(0,x)$ and  $\vrho(1,x)$ can be evaluated by noting that 
$v(u)$ has poles at $u=1$ and $u=-1$, and that integrals can be reduced to residues at the respective poles.\footnote{The branch cuts and poles comprise two copies of those for the BCFTs studied in \cite{He:2024djr} and used in sec.~\ref{sec:BCFT}.} 
The moments at the right boundary are given by 
\begin{align}\label{eq:mom-NS5-gen-right}
M_m&=\int da \trho_{N5}(a) a^{2m} 
=N_5^{2m}\int dx \vrho(1,x) x^{2m}
\nonumber\\
&= \left(\frac{N_5}{2\pi}\right)^{2m+1}\frac{1}{\pi}\Im \int du \frac{dv}{du} \left( N_5 k_0\frac{1-u}{1+u}-N_5 k_1\frac{1+u}{1-u}-i\sum_{p=1}^P N_{\rm D5}^{(p)}\ln\left(\frac{\frac{1-u}{1+u}-i e^{\delta_p}}{\frac{1-u}{1+u}+ie^{\delta_p}}\right) \right)(v-i\pi)^{2m}
\nonumber\\
&=\left(\frac{N_5}{2\pi}\right)^{2m+1}  \Re \Res_{u=-1}{ \frac{dv}{du} \left( \pi K_0\frac{1-u}{1+u}-\pi K_1\frac{1+u}{1-u}-i\sum_{p=1}^P N_{\rm D5}^{(p)}\ln\left(\frac{\frac{1-u}{1+u}-i e^{\delta_p}}{\frac{1-u}{1+u}+ie^{\delta_p}}\right) \right)(v-i\pi)^{2m}}.
\end{align}
At the left boundary, in full analogy,
\begin{align}
M_m=&\int da \trho_{1}(a) a^{2m}
=N_5^{2m}\int dx \vrho(0,x) x^{2m}
\nonumber\\
&= \left(\frac{N_5}{2\pi}\right)^{2m+1}\frac{1}{\pi}\Im \int du \frac{dv}{du} \left( N_5 k_0\frac{1-u}{1+u}-N_5 k_1\frac{1+u}{1-u}-i\sum_{p=1}^P N_{\rm D5}^{(p)}\ln\left(\frac{\frac{1-u}{1+u}-i e^{\delta_p}}{\frac{1-u}{1+u}+ie^{\delta_p}}\right) \right)v^{2m}
\nonumber\\
&=-\left(\frac{N_5}{2\pi}\right)^{2m+1}  \Re \Res_{u=1}{ \frac{dv}{du} \left( \pi K_0\frac{1-u}{1+u}-\pi K_1\frac{1+u}{1-u}-i\sum_{p=1}^P N_{\rm D5}^{(p)}\ln\left(\frac{\frac{1-u}{1+u}-i e^{\delta_p}}{\frac{1-u}{1+u}+ie^{\delta_p}}\right) \right)v^{2m}}~,
\end{align}
where the extra minus sign comes from the opposite orientation of the integration contour.
Then the defect one-point functions are given by \eqref{eq:one-point-gen}.
The results on the left and right are related by the exchange of parameters
\begin{align}\label{eq:k2 k2}
	K_0&\leftrightarrow K_1~,
	&	K_2&\leftrightarrow K_3~,
	&   \delta_p&\leftrightarrow -\delta_p~,
\end{align}
so we focus on the right boundary at $z=1$ in the following.
The auxiliary one-point function can be expressed in a closed form in terms of the residues by \eqref{eq:one-point-int}
\bea\label{eq:one-point-NS5-int-gen}
\langle\hat{\mathcal O}_J\rangle
&=\frac{1}{(-2)^{J/2}\sqrt{J}}\left[N_{\rm D3}\delta_{J,2}+2 N_5 \int dx\, \trho_{N_5}(x)\,T_J\left(\frac{2\pi N_5 x}{\sqrt{\lambda}}\right)\right]\\
&=\frac{1}{(-2)^{J/2}\sqrt{J}}\biggr[N_{\rm D3}\delta_{J,2}+\\
& N_5\Re \Res_{u=1}\left[ \frac{dv}{du} \left( K_0\frac{1-u}{1+u}-K_1\frac{1+u}{1-u}-\frac{i}{\pi}\sum_{p=1}^P N_{\rm D5}^{(p)}\ln\left(\frac{\frac{1-u}{1+u}-i e^{\delta_p}}{\frac{1-u}{1+u}+ie^{\delta_p}}\right) \right)T_J\left(\frac{N_5(v-i\pi)}{\sqrt{\lambda}}\right)\right]\Bigg]~. 
\eea
Then the actual one-point functions can be obtained from \eqref{eq:mixing-sum}.

\subsection{D3/NS5 defect}\label{sec:NS5-defect}

We now specialize the discussion to a defect engineered by a single group of NS5-branes. We take equal gauge groups and couplings in the ambient theories to either side of the defect.
The brane setup is as in fig.~\ref{fig:D3D5-defect} but with the vertical lines denoting NS5 branes.
The NS5-branes then engineer a sequence of $N_5-1$ 3d gauge nodes, with a quiver of the form
\begin{align}\label{eq:NS5-defect-quiver}
	\widehat{U(N_{\rm D3})}-U(N_{\rm D3})-\ldots - U(N_{\rm D3})-\widehat{U(N_{\rm D3})}
\end{align}
The setup is S-dual to the D3/D5 defect discussed in sec.~\ref{sec:D3D5-defect}.
The supergravity dual is obtained by setting $K_3=K_2$, $K_1=K_0$ and $N_{\mathrm D5}=0$ in \eqref{eq: a1-a2-inter-NS5-gen}. The holomorphic functions are then
\begin{align}
	\mathcal A_1&=\frac{\pi \alpha'}{4}K_0\left( e^z-e^{-z}\right)~,
	\nonumber\\
	\mathcal A_2&=\frac{\pi\alpha'}{4}K_2\left( e^z+e^{-z}\right)-\frac{\alpha'}{4}N_5 \ln \tanh\left(\frac{z}{2}\right)         ~.
\end{align}
From the asymptotic expansion of the supergravity fields we find
\begin{align}
	e^{2\phi}&=\frac{K_2}{K_0}= \frac{\gym ^2}{4\pi}~,&
	N_{\rm D3}&=\pi K_2 K_0+ K_0 N_5~. 
\end{align}
This can be inverted to express the supergravity parameters in terms of field theory quantities as
\begin{align}
	K_0&=\frac{4\pi}{\gym^2}K_2~,
	&
	K_2 &= -\frac{N_5}{2 \pi} + \frac{\sqrt{N_5^2 + \gym^2 N_{D3}}}{2 \pi}~.
\end{align}
The eigenvalue density and moments for the 4d node at $z=1$ are obtained as specializations of \eqref{eq: saddle-D3/D5/NS5-gen}, \eqref{eq:mom-NS5-gen-right} as
\begin{align}
\trho_t(a)
&=K_0 \Im\left [ W_{k_2,k_2}(-e^v)- W_{k_2,k_2}(-e^v)^{-1} \right]~, 
&
v&=\frac{2\pi a+i\pi t}{N_5}~,
\nonumber\\
M_m&=\left(\frac{N_5}{2\pi}\right)^{2m+1} \pi K_0 \Re \Res_{u=-1}\left[{\frac{dv}{du} \left(\frac{1-u}{1+u}-\frac{1+u}{1-u}\right)(v-i\pi)^{2m}}\right]~.
\end{align}
The lowest moment reproduces the correct normalization, $M_0=K_0 \left(N_5 + K_2 \pi\right)=N_{\mathrm D3}$.
The auxiliary one-point functions are obtained from \eqref{eq:one-point-NS5-int-gen} as
\bea
\ehmo{J}&=\frac{1}{(-2)^{J/2}\sqrt{J}}\biggr[N_{\rm D3}\delta_{J,2}+N_5 K_0\Re \Res_{u=1}\left[{ \frac{dv}{du} \left(  \frac{1-u}{1+u}-\frac{1+u}{1-u} \right)}T_J\left(\frac{N_5(v-i\pi)}{\sqrt{\lambda}}\right)\right]\biggr]~. 
\eea
To express them more directly in terms of field theory quantities, we express the 5-brane numbers as in (\ref{eq:N5ND5-def}) and define a $d$ similar to that of \eqref{eq:cND5-kappa-def} with vanishing $k$,
\begin{align}
	d&=\sqrt{1 + \mathfrak N_5^2} - \mathfrak N_5~,
	&
	\mathfrak N_5&=\frac{N_5}{\sqrt{\lambda}}~.
\end{align}
Then
\begin{align}
	\ehmo{J}&=\frac{N_{\rm D3}}{(-2)^{J/2}\sqrt{J}}\left[\delta_{J,2}+2\mathfrak N_5 d \Res_{u=1}\left[ \frac{dv}{du} \left(  \frac{1-u}{1+u}-\frac{1+u}{1-u} \right)T_J\left(\mathfrak N_5(v-i\pi)\right)\right]\right]~,
\end{align}
with
\be
v=\frac{d}{2\mathfrak N_5}\left(\frac{1-u}{1+u}+\frac{1+u}{1-u}\right)-\log u~.
\ee
The actual one-point functions are obtained via \eqref{eq:mixing-sum}; the first examples are
\begin{align}\label{eq:NS5-defect-one-point-ex}
	\langle \mathcal O_{J=2}\rangle&=
	\frac{d^4+2 d^2-3}{6 \sqrt{2}}N_{\rm D3}~,
	\qquad\qquad\qquad\qquad
	\langle \mathcal O_{J=4}\rangle=
	\frac{6 d^6-d^4-20 d^2+15}{60}N_{\rm D3}~,
	\nonumber\\
	\langle \mathcal O_{J=6}\rangle&=\frac{190 d^8-344 d^6-371 d^4+1050 d^2-525}{840 \sqrt{6}}N_{\rm D3}~,
		\nonumber\\
	\langle \mathcal O_{J=8}\rangle&=\frac{644 d^{10}-2161 d^8+880 d^6+4312 d^4-5880 d^2+2205}{5040 \sqrt{2}}N_{\rm D3}~.
\end{align}
Since the dCFT is symmetric under reflection across the defect, with identical gauge groups and couplings on the two sides, the one-point functions on the two sides are identical (cf.~\eqref{eq:k2 k2}).

The NS5 defect one-point functions (\ref{eq:NS5-defect-one-point-ex}) can be compared to those for the D3/D5 defect in (\ref{eq:D3D5-defect-ex}). Upon replacing $\mathfrak N_5$ by $\mathfrak N_{\rm D5}$, which implements the action of S-duality, they agree up to an overall factor $(-1)^{J/2}$.
This is the expected relation and provides a non-trivial consistency check.

\subsection{D3/NS5 interface}\label{sec:NS5-interface-one-pt}

The setup of the previous section can be generalized to allow for different gauge groups and couplings on the two sides of the interface, by terminating D3-branes on NS5-branes. 
We take a brane setup of the general form as in fig.~\ref{fig:D3D5-interface}, now with the vertical lines denoting $N_5$ NS5 branes, and take all NS5-branes to have the same number of D3-branes ending on them.
The quiver then takes the form
\begin{align}\label{eq:NS5-interface-quiver}
	\widehat{U(N_{\rm D3}^L)}-U(N_1)-\ldots - U(N_{N_5-1})-\widehat{U(N_{\rm D3}^{\rm R})}
\end{align}
with the ranks of the 3d gauge groups decreasing in steps of $(N_{\rm D3}^L-N_{\rm D3}^{\rm R})/N_5$ from left to right.
These setups are S-dual to the D3/D5 interfaces of sec.~\ref{sec:D3D5-gen}.
The holomorphic functions parametrizing the gravity duals are
\begin{align}
	\mathcal A_1&=\frac{\pi \alpha'}{4}\left(K_0 e^z-K_1 e^{-z}\right)~,
	\nonumber\\
	\mathcal A_2&=\frac{\pi\alpha'}{4}\left(K_2 e^z+K_3 e^{-z}\right)-\frac{\alpha'}{4}N_5 \ln \tanh\left(\frac{z}{2}\right)        ~.
\end{align}
with
\begin{align}
	e^{2\phi_L}&=\frac{K_3}{K_1}~, 
	&N_{\rm D3}^{\rm L}&=\frac{\pi}{2} (K_0K_3+K_1K_2)+K_1 N_{\rm 5}~,
	\nonumber\\
	e^{2\phi_R}&=\frac{K_2}{K_0}~,&
	k\equiv N_{\rm D3}^{\rm L}-N_{\rm D3}^{\rm R}&=N_{\rm 5}(K_1-K_0)~.
\end{align}
This can be inverted to express the gravity parameters in terms of field theory data. To this end, we introduce, similar to (\ref{eq:cND5-kappa-def}),
\begin{align}\label{eq: field to inter NS5}
	\mathfrak N_5 &\equiv \frac{N_5}{\sqrt{\bar g^2 N_{\rm D3}^{\rm R}}}~,
	&
	\kappa &\equiv \frac{k}{4 N_{\rm D3}^{\rm R}\mathfrak N_5}~,
	&
	d &\equiv \sqrt{(\kappa + \mathfrak N_5)^2 + 1} \;-\; \mathfrak N_5~,
\end{align}
where we parametrize the number of NS5-branes by $\mathfrak N_5$ and the difference in ranks by $\kappa$.
We also define the ``average" of the left and right gauge couplings
\begin{align}
	\bar g^2&\equiv\frac{g_L^2 + g_R^2}{2}~.
\end{align}
This is related to the $\bar g$ in D3/D5 (\ref{eq:cND5-kappa-def}) under S-duality as: $ S(\bar g_{\mathrm{there}})=\frac{4\pi}{\bar g_{\mathrm{here}}}$.
We then find as solution with positive $K_i$
\begin{align}\label{eq: K to kappa NS5 gen}
	K_0&=2\left(d-\kappa\right) \sqrt{\frac{N_{\rm D3}^{\rm R}}{\bar g^2}}~,
	&
	K_2&=\frac{g_R^2}{4\pi}K_0~,
	\nonumber\\
	K_1&=2\left(d+\kappa\right) \sqrt{\frac{N_{\rm D3}^{\rm R}}{\bar g^2}}~,
	&
	K_3&=\frac{g_L^2}{4\pi}K_1~.
\end{align} 
We focus on the 4d $U(N_{\rm D3}^{\rm R})$ $\mathcal N=4$ SYM side of the interface.
Again the eigenvalue densities and moments follow from \eqref{eq: saddle-D3/D5/NS5-gen}, \eqref{eq:mom-NS5-gen-right}
\begin{align}
\trho_t(a)
&=\Im\left [K_0 W_{k_2,k_3}(-e^v)-K_1 W_{k_2,k_3}(-e^v)^{-1}
\right]~, 
&
v&=\frac{2\pi a+i\pi t}{N_5}~,
\nonumber\\
M_m&=(\frac{N_5}{2\pi})^{2m+1} \Re \Res_{u=-1}{ \frac{dv}{du} \left( K_0\frac{1-u}{1+u}- K_1\frac{1+u}{1-u}- \right)(v-i\pi)^{2m}}~.
\end{align}
The lowest moment gives the number of semi-infinite D3 branes,
$M_0 = K_0 N_5 + \frac{\pi}{2} \left(K_1 K_2 + K_0 K_3\right)=N_{\mathrm{D3}}^{\rm R}$, reflecting the correct normalization of the eigenvalue density.
The auxiliary one-point functions follow from \eqref{eq:one-point-NS5-int-gen},
\begin{align}
	\langle\hat{\mathcal O}_J\rangle&=\frac{N_{\rm D3}^{\rm R}}{(-2)^{J/2}\sqrt{J}}\biggr[\delta_{J,2}+ 2\mathfrak{N}_5\Res_{u=1}\left[\frac{dv}{du} \left(  (d-\kappa)\frac{1-u}{1+u}-(d+\kappa)\frac{1+u}{1-u}\right)T_J\left(\frac{N_5(v-i\pi)}{\sqrt{\lambda}}\right)\right]\biggr]~,
\nonumber\\
v&=g_R^2\frac{d-\kappa}{2 \bar g^2 \mathfrak N_5} \frac{1-u}{1+u}+g_L^2\frac{d+\kappa}{2 \bar g^2 \mathfrak N_5}\frac{1+u}{1-u}-\log u~.
\end{align}
We give the first examples of the actual one-point functions following from (\ref{eq:mixing-sum}) for the special case of equal gauge couplings, $g_L=g_R$; this allows for direct comparison with the S-dual case in \eqref{eq:D3D5gen-ex} upon replacing $\mathfrak N_5$ by $\mathfrak N_{\rm D5}$. We find
\bea \label{eq: onepoin D3NS5gen}
\langle \mathcal{O}_{J=2} \rangle &= 
- \frac{\sqrt{2} (d - \kappa) (2d + 3\mathfrak{N}_5 + \kappa)}{3 (d + 2\mathfrak{N}_5 + \kappa)}N_{\mathrm{D3}}^{\mathrm{R}} \mathfrak{N}_5~, \\
\langle \mathcal{O}_{J=4} \rangle &= 
\frac{ (d - \kappa) (4d^2 + 50d\mathfrak{N}_5 + 60\mathfrak{N}_5^2 - 3d\kappa + 10\mathfrak{N}_5\kappa - \kappa^2)}{30 (d + 2\mathfrak{N}_5 + \kappa)^2}N_{\mathrm{D3}}^{\mathrm{R}} \mathfrak{N}_5~. \\
\eea
Setting $\kappa=0$ and expressing $\mathfrak{N}_5$ in terms of $d$, we recover the defect one point functions \eqref{eq:NS5-defect-one-point-ex}. Again the one point functions on the two sides of the interfaces are related by the symmetry \eqref{eq:k2 k2}; in terms of field theory parameters this now amounts to
\begin{align}\label{eq: lr D3NS5 gen}
	g_L&\leftrightarrow g_R~,
	&	\kappa&\leftrightarrow -	\kappa~.
\end{align}
From \eqref{eq: K to kappa NS5 gen} it is manifest that this operation exchanges the $K_i$'s between the two sides.

\let\oldaddcontentsline\addcontentsline
\renewcommand{\addcontentsline}[3]{}
\begin{acknowledgments}
	DH is supported by FWO-Vlaanderen project G012222N and by the VUB Strategic Research Program High-Energy Physics.
\end{acknowledgments}
\let\addcontentsline\oldaddcontentsline

\appendix
\section{Normal ordering}\label{sec:app-diag}
As is discussed in sec. \ref{sec:one-point-functions}, we need to diagonalize the two point functions including multi-trace contributions, i.e.\ with finite $N$ effects.
An algorithm to do this is formulated in \cite{Billo:2017glv}.
We briefly review it in this appendix and give explicit examples.

The way to do the diagonalization is to subtract self-contractions. The operators we study should not have self-contractions in the 4d theory without defects or boundaries. 
However, results from insertions in the matrix models do not enjoy this property so we need to subtract them manually. 
Given an operator $\mo(a)$ with conformal dimension $\Delta$, we need to make it orthogonal to all the lower dimensional operators $\mo_p(a)$. 
This can be realized by the following normal ordering
\begin{equation}
	: \mo(a): =O(a)-\sum_{p, q}\left\langle O(a) O_p(a)\right\rangle_0 C^{p q} O_q(a)~,
\end{equation}
where elements of the matrix $C$ are the two point functions of the lower dimensional operators
\begin{equation}
	C_{p q}=\left\langle O_p(a) O_q(a)\right\rangle_0~.
\end{equation}
The subscript $0$ of the correlators stands for pure 4d theories without boundaries or defects.
In our matrix model method we are concerned about operators of the form $\tr a^J$ for even $J$, and the lower dimensional operators can be both single trace and multi trace operators.
What remains is to determine the matrix $C$. For convenience we introduce the notation for correlation functions
\begin{equation}
	t_{n_1, n_2, \ldots} \equiv\left\langle\operatorname{tr} a^{n_1} \operatorname{tr} a^{n_2} \ldots\right\rangle_0~,
\end{equation}
which can be computed directly since the partition function can be written as a Gaussian. The result is a recursive representation with initial conditions for single trace correlators,
\begin{align}
	t_0 &= N, & t_1&=0, & t_2&=\frac
	{N^2-1}{2}~, 
\end{align}
and reduction conditions for multi trace correlators
\begin{equation}
	\begin{aligned}
		& t_{0, n_1, n_2, \ldots}=N t_{n_1, n_2, \ldots} ~,\\
		& t_{1, n_1, n_2, \ldots}=0~, \\
		& t_{2, n_1, n_2, \ldots}=\frac{N^2-1+n_1+n_2+\ldots}{2} t_{n_1, n_2, \ldots}~.
	\end{aligned}
\end{equation}
Then general $t$s with indices greater than $2$ are given by
\begin{equation}
	\begin{aligned}
		t_n= & \frac{1}{2} \sum_{m=0}^{n-2}\left(t_{m, n-m-2}-\frac{1}{N} t_{n-2}\right)~, \\
		t_{n, n_1}= & \frac{1}{2} \sum_{m=0}^{n-2}\left(t_{m, n-m-2, n_1}-\frac{1}{N} t_{n-2, n_1}\right)+\frac{n_1}{2}\left(t_{n+n_1-2}-\frac{1}{N} t_{n-1, n_1-1}\right)~, \\
		t_{n, n_1, n_2}= & \frac{1}{2} \sum_{m=0}^{n-2}\left(t_{m, n-m-2, n_1, n_2}-\frac{1}{N} t_{n-2, n_1, n_2}\right)+\frac{n_1}{2}\left(t_{n+n_1-2, n_2}-\frac{1}{N} t_{n-1, n_1-1, n_2}\right)~, \\
		& \quad+\frac{n_2}{2}\left(t_{n+n_2-2, n_1}-\frac{1}{N} t_{n-1, n_1, n_2-1}\right)~,\\
		& \ldots~.
	\end{aligned}
\end{equation}
We take $\tr a^4$ and $\tr a^6$ as examples for the subtraction. To get the normal ordered version, for $\tr a^4$ we need to subtract its contraction with the identity and $\tr a^2$, while for $\tr a^6$ the lower dimensional operators are the identity, $\tr a^2$, $\tr a^4$ and the double trace operator $(\tr a^2)^2$. The results to the leading order in $N$ up to $:\tr a^8 :$ are
\begin{equation}\label{eq: normal tr a}
	\begin{aligned}
		& : \tr a^4:=\tr a^4-2N \tr a^2+\frac{1}{2}N^3,~ \\
		& : \tr a^6:=\tr a^6 
		- 3 N \tr a^4 
		- \frac{3}{2} (\tr a^2)^2 
		+ \frac{15}{4} N^2 \tr a^2 
		- \frac{5 N^4}{8}~,\\
		& : \tr a^8:=\tr a^8 
		- 4 N \tr a^6 
		+ 7 N^2 \tr a^4 
		- 4 \tr a^2 \tr a^4 
		+ 7 N \, (\tr a^2)^2 
		- 7 N^3 \tr a^2 
		+ \frac{7 N^5}{8}. 
	\end{aligned}
\end{equation}
Now we give explicit procedure to get normal ordered $\mo_2$, $\mo_4$, $\mo_6$ and $\mo_8$. From \eqref{eq:OJ-insert}, we have the infinite $N$ results
\bea \label{eq: O from cheby}
\hat{\mo}_2 & = -\frac{4\sqrt{2} \, \tr a^2 \, \pi^2}{\lambda} + \frac{N}{2\sqrt{2}}, \\
\hat{\mo}_4 & = \frac{32 \, \tr a^4 \, \pi^4}{\lambda^2} - \frac{8 \, \tr a^2 \, \pi^2}{\lambda} + \frac{N}{4}, \\
\hat{\mo}_6 & = -\frac{256 \sqrt{\frac{2}{3}} \, \tr a^6 \, \pi^6}{\lambda^3} + \frac{32 \sqrt{6} \, \tr a^4 \, \pi^4}{\lambda^2} - \frac{3 \sqrt{6} \, \tr a^2 \, \pi^2}{\lambda} + \frac{N}{4\sqrt{6}}, \\
\hat{\mo}_8 & = \frac{1024 \sqrt{2} \, \tr a^8 \, \pi^8}{\lambda^4} - \frac{512 \sqrt{2} \, \tr a^6 \, \pi^6}{\lambda^3} + \frac{80 \sqrt{2} \, \tr a^4 \, \pi^4}{\lambda^2} - \frac{4 \sqrt{2} \, \tr a^2 \, \pi^2}{\lambda} + \frac{N}{16\sqrt{2}}.
\eea
Now using \eqref{eq: normal tr a} we compute the leading-order finite $N$ results $\mo_J(a)=\frac{1}{(- N)^\frac{J}{2} \sqrt{J}}:\tr (\sqrt{\frac{8\pi^2 N}{\lambda}}a)^J :$, 
\bea\label{eq: O from finite}
\mo_2 & = -\frac{4\sqrt{2} \, \tr a^2 \, \pi^2}{\lambda} + \frac{N}{2\sqrt{2}}, \\
\mo_4 & = \frac{32 \, \tr a^4 \, \pi^4}{\lambda^2} - \frac{8 \, \tr a^2 \, \pi^2}{\lambda} + \frac{N}{4}, \\
\mo_6 & = -\frac{256 \sqrt{2/3} \, \tr a^6 \, \pi^6}{\lambda^3} 
+ \frac{32 \sqrt{6} \, \tr a^4 \, \pi^4}{\lambda^2} 
- \frac{5 \sqrt{6} \, \tr a^2 \, \pi^2}{\lambda} 
+ \frac{16\sqrt{6} \, (\tr a^2)^2}{6 N} 
+ \frac{5 N}{8\sqrt{6}}\,,
\\
\mo_8 & = \frac{7 N}{16 \sqrt{2}} 
+ \frac{1024 \sqrt{2} \, \tr a^8 \, \pi^8}{\lambda^4} 
- \frac{512 \sqrt{2} \, \tr a^6 \, \pi^6}{\lambda^3}
- \frac{512 \sqrt{2} \, \tr a^2 \, \tr a^4 \, \pi^6}{N \lambda^3} \\
&\hphantom{=}+ \frac{112 \sqrt{2} \, \tr a^4 \, \pi^4}{\lambda^2} 
+ \frac{112 \sqrt{2} \, (\tr a^2)^2 \, \pi^4}{N \lambda^2} 
- \frac{14 \sqrt{2} \, \tr a^2 \, \pi^2}{\lambda}\,.       
\eea
Comparing \eqref{eq: O from cheby} and  \eqref{eq: O from finite}, one can derive diagonalized operators $\mo_J$ in terms of $\hat{\mo}_J$. This leads to (\ref{eq:op-mixing}) as first examples.

\bibliography{BCFT}
\end{document}